\documentclass[showpacs,amssymb,preprint,preprintnumbers,nofootinbib,superscriptaddress]{revtex4}
\usepackage{amsmath}
\usepackage{graphicx}
\usepackage{latexsym}
\usepackage{amsfonts}
\usepackage{url,hyperref}
\usepackage{bm}
\usepackage{textcomp}
\usepackage{color}
\usepackage{bbm}
\usepackage{slashed}

\newcommand{\beq}{\begin{equation}}
\newcommand{\eeq}{\end{equation}}
\def\bea{\begin{eqnarray}}
\def\eea{\end{eqnarray}}

\usepackage{caption}
\usepackage{epstopdf}
\usepackage{subcaption}
\usepackage{placeins}
\renewcommand{\theequation}{\arabic{section}.\arabic{equation}}

\begin{document}
\hfill{WITS-MITP-022}
\title{Spin-3/2 fields in $D$-dimensional Schwarzschild black hole space-times}
\author{C.-H. Chen}
\email[Email: ]{chunhungchen928@gmail.com}
\affiliation{Department of Physics, Tamkang University, Tamsui, Taipei, Taiwan}
\author{H.~T.~Cho}
\email[Email: ]{htcho@mail.tku.edu.tw}
\affiliation{Department of Physics, Tamkang University, Tamsui, Taipei, Taiwan}
\author{A.~S.~Cornell}
\email[Email: ]{alan.cornell@wits.ac.za}
\affiliation{National Institute for Theoretical Physics; School of Physics, University of the
Witwatersrand, Wits 2050, South Africa}
\author{G. Harmsen}
\email[Email: ]{gerhard.harmsen5@gmail.com}
\affiliation{National Institute for Theoretical Physics; School of Physics, University of the
Witwatersrand, Wits 2050, South Africa}

\begin{abstract}
\par In previous works we have studied spin-3/2 fields near 4-dimensional Schwarzschild black holes. The techniques we developed in that case have now been extended here to show that it is possible to determine the potential of spin-3/2 fields near $D$-dimensional black holes by exploiting the radial symmetry of the system. This removes the need to use the Newman-Penrose formalism, which is difficult to extend to $D$-dimensional space-times. In this paper we will derive a general $D$-dimensional gauge invariant effective potential for spin-3/2 fields near black hole systems. We then use this potential  to determine the quasi-normal modes and absorption probabilities of spin-3/2 fields near a $D$-dimensional Schwarzschild black hole.
\end{abstract}

\date{\today}
\maketitle


\section{Introduction}\label{Intro}\setcounter{equation}{0}
\par With the introduction of supergravity theories there has been a lot of interest in understanding the gravitino, the supersymmetric partner of the graviton. This particle is predicted to be a spin-3/2 particle and would behave like a Rarita-Schwinger field. In many supergravity theories these fields act as a sources of torsion and curvature of the space-time \cite{das1976gauge,grisaru1977supergravity}. It is also predicted that the gravitino is the lighest or second lightest supersymmetric particle. This makes the gravitino an ideal candidate for studying supersymmetric gravitational theories. Much of the research into gravitinos has been focused on the particle and not its interaction with curved space-time, specifically near black holes. In previous works we have investigated this space-time interaction for 4-dimensional Schwarzschild black holes with a spin-3/2 field \cite{chen2015gravitino}. We did this by exploiting the radial symmetry of our system, which allowed us to separate the metric into a radial-time part and an angular part. We could then calculate the eigenvalues of our eigenspinor-vectors in the angular part by using the works of Camporesi \textit{et al.} \cite{camporesi1996eigenfunctions}. Relating the eigenvalues of our radial-time part to those for the angular part we could easily determine the effective potential for our fields near the black hole. Using the same approach we hope to be able to investigate space-times with dimensions greater than 4.
\par This paper aims to derive a gauge invariant $D$-dimensional effective potential for spin-3/2 fields near Schwarzschild black holes. Using this potential we can study the evolution of our spin-3/2 particles as they propagate through the curved space-time. This evolution through space-time is characterized by an oscillation of the space-time. Near black holes these oscillations have a single frequency with a damping term, and are called Quasi-Normal Modes (QNMs). The QNMs characterize the parameters of the black hole \cite{konoplya2011quasinormal}. A variety of numerical and semi-analytic techniques have been used to determine the numerical values for the emitted QNMs \cite{kokkotas1999quasi, berti2009quasinormal}. We will use the WKB method and a method developed by some of us called the Improved Asymptotic Iterative Method (Improved AIM) to calculate these values \cite{cho2012new}.
\par As we have done previously we will also look at the absorption probabilities of our spin-3/2 particles, as this will give us an insight into the grey body factors and emission cross-section of our black holes. These are required in order to understand the stability and evolution of our black holes.
\par The paper will be structured as follows. In the next section we determine the eigenvalues for the spinor-vectors on an $N$-sphere. In Sec.\ref{Sec:Potential} we calculate the potential functions for spin-3/2 fields in this space-time. We then use the potential to determine the QNMs and absorption potential associated to our spin-3/2 particles, the results are given in Secs.\ref{Sec:QNMS} and \ref{Sec:Absorption}. Final discussions and conclusions are given in Sec.\ref{Sec:conclusions}.


\section{Spinor-vector eigenmodes on $S^{N}$}\setcounter{equation}{0}
\label{sec:Eigenmodes}
\subsection{The $N$-sphere}
\par The metric that describes the angular part of our space-times is simply the metric for a sphere. The metric for the $N$-sphere, $S^{N}$, is given as
\begin{equation}
\begin{aligned}
d\Omega_{N}^{2} = \sin^{2}\theta_{N}d\widetilde{\Omega}_{N-1}^{2} + d\theta_{N}^{2} \: ,\\
\end{aligned}
\end{equation}
where $d\widetilde{\Omega}_{N-1}$ is the metric of $S^{N-1}$. In the rest of this section we will denote terms for $S^{N-1}$ with tildes. Non-zero Christoffel symbols for the $S^{N-1}$ are
\begin{equation}\label{Eq:SphereChristofel}
\begin{aligned}
&\Gamma^{\theta_{N}}_{\theta_{i}\theta_{j}} = - \sin\theta_{N}\cos\theta_{N}\widetilde{g}_{\theta_{i}\theta_{j}} \: ; \: \: \Gamma^{\theta_{j}}_{\theta_{i}\theta_{N}} = \cot\theta_{N}\widetilde{g}^{\theta_{j}}_{\theta_{i}} \: ; \: \: \Gamma^{\theta_{k}}_{\theta_{i}\theta_{j}} = \widetilde{\Gamma}^{\theta_{k}}_{\theta_{i}\theta_{j}} \: .
\end{aligned}
\end{equation}

In order to determine our covariant derivatives on $S^{N}$ we need to determine the appropriate spin connections. We will use the $n$-bein formalism in order to relate components on our curved space to those of an orthonormal basis \cite{parker2009quantum}. Our metric is related as follows,
\begin{equation}
\begin{aligned}
&g_{\mu \nu} = e^{a}_{\mu}e^{b}_{\nu}\delta_{ab}\: ,\: \: e^{a}_{\mu}e^{\mu}_{b} = \delta^{a}_{b} \: , \: \: e^{a}_{\mu}e^{\nu}_{a} = \delta^{\nu}_{\mu} \: ,
\end{aligned}
\end{equation}
where Greek letters represent our world indices and Latin letters represent our Lorentz indices. For $S^{N}$ the $n$-bein is given as,
\begin{equation}\label{VielBeinNSphere}
\begin{aligned}
e^{\theta_{N}}_{N} & = 1 \: , \: \: e^{\theta_{i}}_{i} & = \frac{1}{\sin(\theta)} \widetilde{e}^{\theta_{i}}_{i} \: .  \\
\end{aligned}
\end{equation}
We can relate gamma matrices on the orthogonal basis to gamma matrices of those on $S^{N}$ as follows,
\begin{equation}
\begin{aligned}
&\gamma^{\theta_{i}}=e^{\theta_{i}}_{i}\gamma^{i} = \frac{1}{\sin\theta_{N}}\widetilde{e}^{\theta_{i}}_{i}\gamma^{i} \: , \: \: \gamma^{\theta_{N}}=e^{\theta_{N}}_{N}\gamma^{N}=\gamma^{N} ,
\end{aligned}
\end{equation}
where the gamma matrices on the orthogonal basis obey the Clifford algebra. Spin connections are calulated as \cite{parker2009quantum}
\begin{equation}
\begin{aligned}
\omega_{\mu a b} = e^{\alpha}_{a}\left(\partial_{\mu}e_{\alpha b} - \Gamma^{\rho}_{\mu \alpha}e_{\rho b} \right).
\end{aligned}
\end{equation}
Non-zero spin connections on $S^{N}$ are then determined to be

\begin{equation}
\begin{aligned}
\omega_{\theta_{i} j N} = \cos\theta_{N}\widetilde{e}_{\theta_{i}j} \: ,\: \: \omega_{\theta_{i} j k} = \widetilde{\omega}_{\theta_{i} j k} \: .
\end{aligned}
\end{equation}
The covariant derivative for the spinor-vector field is
\begin{equation}
\begin{aligned}
\nabla_{\mu}\psi_{\nu}=\partial_{\mu}\psi_{\nu}-\Gamma^{\rho}_{\mu\nu}\psi_{\rho}+\omega_{\mu}\psi_{\nu} \: ,
\end{aligned}
\end{equation}
where
\begin{equation}
\begin{aligned}
\omega_{\mu}=\frac{1}{2}\omega_{\mu a b}\Sigma^{a b}\ \ , \ \ \Sigma^{a b}=\frac{1}{4}[\gamma^{a},\gamma^{b}] \: .
\end{aligned}
\end{equation}
\par We can now determine the eigenvalues for our spinors and spinor-vectors on $S^{N}$. The eigenvalues for our spinors have already been determined by Camporesis and Higuchi \cite{camporesi1996eigenfunctions}, a brief overview is given in Appendix \ref{Apend:NSphere}. We use the eigenvalues of our spinors to determine the eigenvalues for our spinor-vectors.
\par We denote the spinor-vectors as $\psi_{\mu}$, where each of the components are spinors. To begin our investigation into spinor-vectors we find two orthogonal eigenspinor-vectors on $S^{2}$, which can be written as linear combinations of the basis $\gamma_{\mu}\psi_{(\lambda)}$ and $\nabla_{\mu}\psi_{(\lambda)}$, where $\psi_{(\lambda)}$ is the eigenspinor on $S^{2}$. Note that we use $(\lambda)$ to highlight that $\lambda$ is not an index. These are ``non Transverse and Traceless eigenmodes" (non-TT modes), as they do not satisfy the transverse and traceless conditions. These two eigenspinor-vectors can be generalised to the $S^{N}$ case, and are analogous to the longitudinal eigenmode for vector fields on spheres. These ``non-TT eigenmodes" form a complete set of eigenmodes on $S^{2}$. For higher dimensional surfaces these modes do not represent a complete set and we must introduce TT eigenmodes. We are also required to consider the behavior of the $S^{N}$ spinor-vector components on $S^{N-1}$. Consider the surface $S^{3}$, we expect the following spinor-vector $\psi_{\theta_{i}} = (\psi_{\theta_{3}},\psi_{\theta_{2}},\psi_{\theta_{1}})$, with $\psi_{\theta_{1}},\psi_{\theta_{2}}$ and $\psi_{\theta_{3}}$ representing spinors. Furthermore on $S^{3}$ we expect our ``TT components", $\psi_{\theta_{1}}$ and $\psi_{\theta_{2}}$, to behave like spinor-vectors on $S^{2}$. Since $S^{2}$ only has ``non-TT eigenmodes" we should represent $\psi_{\theta_{1}}$ and $\psi_{\theta_{2}}$ as linear combinations of ``non-TT eigenmodes" on $S^{3}$. As such $\psi_{\theta_{3}}$ acts like a spinor on $S^{2}$, and we represent it using a linear combination of spinor eigenmodes on $S^{2}$. This gives us our first type of ``TT-eigenmode" which we can call the ``TT-mode I".  The complete set of spinor-vectors on $S^{3}$ is therefore given by two ``non-TT eigenmodes" and one ``TT-eigenmode I". On $S^{4}$ spinors $\psi_{\theta_{1}}$, $\psi_{\theta_{2}}$ and $\psi_{\theta_{3}}$ behave like spinor-vectors and $\psi_{\theta_{4}}$ behaves like a spinor on $S^{3}$. We therefore have two types of spinor-vectors on $S^{3}$, which can be represented in two ways. Firstly we can represent $\psi_{\theta_{1}}$, $\psi_{\theta_{2}}$ and $\psi_{\theta_{3}}$ as linear combination of ``non-TT eigenmodes" on $S^{3}$, and $\psi_{\theta_{4}}$ represented with a linear combinations of spinor eigenmodes on $S^{3}$, this is the ``TT-eigenmode I" on $S^{4}$. We could also represent $\psi_{\theta_{1}}$, $\psi_{\theta_{2}}$ and $\psi_{\theta_{3}}$ as ``TT-eigenmodes" on $S^{3}$. Since they are already the ``TT eigenmodes" $\psi_{\theta_{4}}$ must go to zero. In this case we call it the ``TT eigenmode II". Hence the complete set of eigenmodes on $S^{4}$ is given by two ``non-TT eigenmodes", one ``TT eigenmode I" and one ``TT eigenmode II".
\par Generally eigenmodes on $S^{N}$ are represented by two ``non-TT eigenmodes", one ``TT-eigenmode I" and $N-3$ ``TT-eigenmode II'' when $N > 2$. We can now determine the eigenvalues for our spinor-vectors. In the following section we will denote values relating to the surface $S^{N-1}$ with tildes.

\subsection{Spinor-vector non-TT eigenmodes on $S^{N}$}

We denote eigenvalues for the non-TT eigenmode spinor-vectors as $i\xi$. The eigenvalue equation for our eigenspinor-vectors with the Dirac operator is

\begin{equation}\label{Eq:SpinorDirac}
\begin{aligned}
\gamma^{\mu}\nabla_{\mu}\psi_{\nu} = i\xi\psi_{\nu}\: .
\end{aligned}
\end{equation}
We can construct eigenspinor-vectors on $S^{N}$ using the following linear combination
\begin{equation}\label{Eq:SpinorLin}
\begin{aligned}
\psi_{\nu} = \nabla_{\nu}\psi_{(\lambda)} + a \gamma_{\nu}\psi_{(\lambda)}\: ,
\end{aligned}
\end{equation}
where $\psi_{(\lambda)}$ is an eigenspinor on $S^{N}$. Plugging Eq.(\ref{Eq:SpinorLin}) into Eq.(\ref{Eq:SpinorDirac}) we have
\begin{equation}\label{Eq:SpinorDiracExpand}
\begin{aligned}
i\xi \psi_{\nu} =\gamma^{\mu}[\nabla_{\mu},\nabla_{\nu}]\psi_{(\lambda)} + \nabla_{\nu}(\gamma^{\mu}\nabla_{\mu}\psi_{(\lambda)})+ a\{\gamma^{\mu},\gamma_{\nu} \}\nabla_{\mu}\psi_{(\lambda)} - a \gamma_{\nu}(\gamma^{\mu}\nabla_{\mu}\psi_{(\lambda)})\: .
\end{aligned}
\end{equation}
The commutator can be rewritten in terms of the Riemann curvature tensor $R_{\mu \nu }^{\: \: \: \: \: \sigma \rho}$ as,
\begin{equation}
\begin{aligned}
\gamma^{\mu}[\nabla_{\mu},\nabla_{\nu}]\psi_{(\lambda)} = \frac{1}{8}R_{\mu \nu \; \;}^{\: \: \: \:\sigma \rho}[\gamma_{\sigma},\gamma_{\rho}]\psi_{(\lambda)}
= \frac{1}{2}(N-1)\gamma_{\nu}\psi_{(\lambda)}\: .
\end{aligned}
\end{equation}
Then Eq.(\ref{Eq:SpinorDiracExpand}) becomes
\begin{equation}
\begin{aligned}
\gamma^{\mu}\nabla_{\mu}\psi_{\nu} = (i\lambda + 2 a)\left(\nabla_{\nu}\psi_{(\lambda)} + \frac{-i\lambda a + \frac{1}{2}(N-1)}{i\lambda + 2a}\gamma_{\nu}\psi_{(\lambda)} \right) ,
\end{aligned}
\end{equation}
where $i\lambda$ is the spinor eigenvalue on $S^{N}$ Eq.(\ref{Eig(N)}). Comparing with Eq.(\ref{Eq:SpinorLin}) we can show that the non-TT eigenvalues are
\begin{equation}
\begin{aligned}
i\xi = i \lambda + 2a = \pm i \sqrt{j^{2}+(N-1)j+\frac{1}{4}(N-5)(N-1)}\ , \  j=\frac{1}{2}, \frac{3}{2}, \frac{5}{2},...  .
\end{aligned}
\end{equation}
We can write the non-TT eigenspinor-vectors as
\begin{equation}\label{Eq:NonTTEigenVectors}
\begin{aligned}
\psi^{(1)}_{\nu} = & \nabla_{\nu}\psi_{(\lambda)} + \frac{1}{2}\left(- i\lambda + \sqrt{(N-1)-\lambda^{2}} \right)\gamma_{\nu}\psi_{(\lambda)} \: ,\\
\psi^{(2)}_{\nu} = & \nabla_{\nu}\psi_{(\lambda)} + \frac{1}{2}\left(-i\lambda-\sqrt{(N-1)-\lambda^{2}} \right)\gamma_{\nu}\psi_{(\lambda)}\: .
\end{aligned}
\end{equation}

\subsection{Spinor-vector TT eigenmode I on $S^{N}$}

Here we denote the eigenvalues as $i\zeta$, to distinguish them from the non-TT eigenmodes. So our spinor-vector equation with Dirac operator is

\begin{equation}\label{TTDirac}
\begin{aligned}
\gamma^{\mu}\nabla_{\mu}\psi_{\nu} = i\zeta\psi_{\nu}\: .
\end{aligned}
\end{equation}
The transverse traceless condition is
\begin{equation}\label{Eq:TTcond.}
\begin{aligned}
\nabla^{\mu}\psi_{\mu}= \gamma^{\mu}\psi_{\mu} = 0\: .
\end{aligned}
\end{equation}
With the spinor eigenmodes noted in appendix \ref{Apend:NSphere}, we will separate out our lower dimensional part and separately consider the $N$ even and the $N$ odd cases.

\subsubsection{N odd}
Using Eq.(\ref{EQ:GammaOdd})and Eq.(\ref{VielBeinNSphere}) we find that Eq.(\ref{Eq:TTcond.}) becomes
\begin{equation}\label{Eq:TTcondt.1}
\begin{aligned}
\gamma^{\mu}\psi_{\mu} = 0 \implies \psi_{\theta_{N}} = - \frac{1}{\sin\theta_{N}}\gamma^{N}\widetilde{\gamma}^{\theta_{i}}\psi_{\theta_{i}}\: .
\end{aligned}
\end{equation}
Using the Christoffel symbols in Eq.(\ref{Eq:SphereChristofel}) Eq.(\ref{TTDirac}) and Eq.(\ref{Eq:TTcond.}) become
\begin{equation}\label{Eq:TTcondt.2}
\begin{aligned}
\nabla^{\mu}\psi_{\mu} = 0 \implies \left(\partial_{\theta_{N}} + \left(N -\frac{1}{2} \right)\cot\theta_{N} \right)\psi_{\theta_{N}} = - \frac{1}{\sin^{2}\theta_{N}}\widetilde{\nabla}^{\theta_{i}}\psi_{\theta_{i}}\: ,
\end{aligned}
\end{equation}
\begin{equation}\label{Eq:TTcondt.3}
\begin{aligned}
\gamma^{\mu}\nabla_{\mu}\psi_{\theta_{N}} = i \zeta \psi_{\theta_{N}} \implies \gamma^{N}\left(\partial_{\theta_{N}} + \left(\frac{N+1}{2} \right)\cot\theta_{N} \right)\psi_{\theta_{N}} + \frac{1}{\sin\theta_{N}}\widetilde{\gamma}^{\theta_{i}}\widetilde{\nabla}_{\theta_{i}}\psi_{\theta_{N}} = i \zeta \psi_{\theta_{N}}\: ,
\end{aligned}
\end{equation}
\begin{equation}\label{Eq:TTDirac1}
\begin{aligned}
\gamma^{\mu}\nabla_{\mu}\psi_{\theta_{i}} = i \zeta \psi_{\theta_{i}} \implies & \gamma^{N}\left(\partial_{\theta_{N}} + \left(\frac{N-1}{2} \right)\cot\theta_{N} \right)\psi_{\theta_{i}} + 2\cot\theta_{N}\gamma^{N}\widetilde{\Sigma}_{\theta_{i}}^{\theta_{j}}\psi_{\theta_{j}}\\
& \hspace{4cm}+  \frac{1}{\sin\theta_{N}}\widetilde{\gamma}^{\theta_{j}}\widetilde{\nabla}_{\theta_{j}}\psi_{\theta_{i}}=i\zeta\psi_{\theta_{i}}\: .
\end{aligned}
\end{equation}
$\psi_{\theta_{N}}$ behaves like a spinor and we write it as a linear combination of the the eigenspinors on $S^{N-1}$. The $\psi_{\theta_{i}}$ terms behave like spinor-vectors and we write them in terms of ``non-TT mode" eigenspinor-vectors on $S^{N-1}$,
\begin{equation}\label{Eq:PsiNTT}
\begin{aligned}
\psi_{\theta_{N}} = \frac{1}{\sqrt{2}}(1 + i\gamma^{N})A^{(1)}\widetilde{\psi}_{(\lambda)} + \frac{1}{\sqrt{2}}(1-i\gamma^{N})A^{(2)}\widetilde{\psi}_{(\lambda)}\: ,
\end{aligned}
\end{equation}
\begin{equation}\label{Eq:PsiITT}
\begin{aligned}
\psi_{\theta_{i}} = \frac{1}{\sqrt{2}}(1 + i \gamma^{N})(C^{(1)}\widetilde{\nabla}_{\theta_{i}}\widetilde{\psi}_{(\lambda)} + D^{(1)}\widetilde{\gamma}_{\theta_{i}}\widetilde{\psi}_{(\lambda)}) + \frac{1}{\sqrt{2}}(1 - i \gamma^{N})\left(C^{(2)}\widetilde{\nabla}_{\theta_{i}}\widetilde{\psi}_{(\lambda)} + D^{(2)}\widetilde{\gamma}_{\theta_{i}}\widetilde{\psi}_{(\lambda)} \right) .
\end{aligned}
\end{equation}
The coefficients $A^{(1,2)},C^{(1,2)}$ and $D^{(1,2)}$ are functions of $\theta_{N}$ only, and $\widetilde{\psi}_{(\lambda)}$ is the spinor eigenmode on $S^{N-1}$. Using these two definitions we can derive the eigenspinor-vector equations Eqs.(\ref{Eq:TTcondt.1}) to (\ref{Eq:TTDirac1}) by setting $A^{(1,2)} = \left(\sin\theta_{N} \right)^{-\frac{N+1}{2}}\mathbbm{A}^{(1,2)}$. The coefficients of the eigenspinor-vector are
\begin{equation}
\begin{aligned}
\mathbbm{A}^{(1,2)}=\left(\sin\frac{\theta_{N}}{2}\right)^{\left|\frac{1}{2}\pm \widetilde{\lambda}\right| + \frac{1}{2}}\left(\cos\frac{\theta_{N}}{2} \right)^{\left|\frac{1}{2} \mp \widetilde{\lambda} \right|+ \frac{1}{2}}P_{n}^{\left|\frac{1}{2}\pm \widetilde{\lambda}\right|,\left|\frac{1}{2} \mp \widetilde{\lambda} \right|}(\cos\theta_{N})\: ,
\end{aligned}
\end{equation}
\begin{equation}
\begin{aligned}
C^{(1,2)} = & \frac{\sin\theta_{N}}{\widetilde{\lambda}^{2}-\frac{1}{4}\left(N-1 \right)^{2}}\left(\frac{N-1}{2}\cos\theta_{N} \mp \widetilde{\lambda} \right)A^{(1,2)}\\
& \mp \frac{N-1}{N-2}\frac{\zeta\sin^{2}\theta_{N}}{\widetilde{\lambda}-\frac{1}{4}\left(N-1 \right)^{2}}A^{(2,1)}\: ,
\end{aligned}
\end{equation}
\begin{equation}
\begin{aligned}
D^{(1,2)} = & - \frac{i\sin\theta_{N}}{\widetilde{\lambda}^{2} - \frac{1}{4}\left(N-1 \right)^{2}}\left(\widetilde{\lambda}\cos\theta_{N} \mp \frac{N-1}{2} \right)A^{(1,2)}\\
& \pm \frac{i \zeta \widetilde{\lambda}\sin^{2}\theta_{N}}{(N-2)\left(\widetilde{\lambda}^{2} - \frac{1}{4}(N-1)^{2} \right)}A^{(2,1)}\: ,
\end{aligned}
\end{equation}
where $P_{n}^{\left|\frac{1}{2}\pm \widetilde{\lambda}\right|,\left|\frac{1}{2} \mp \widetilde{\lambda} \right|}(\cos\theta_{N})$ is the Jacobi polynomial. The eigenvalues are
\begin{equation}\label{Eq:ZetaSoln}
\begin{aligned}
i\zeta = \pm i\left(n + \left|\widetilde{\lambda} \right| + \frac{1}{2} \right),
\end{aligned}
\end{equation}
where $\left|\widetilde{\lambda} \right| = \widetilde{n} + (N-1)/2$, $\widetilde{n} = 0,\: 1,\:  2,\: ...$ and $n = 0,\: 1,\: 2, \:...$. Eq.(\ref{Eq:ZetaSoln}) can be rewritten as
\begin{equation}\label{Eq:nonTTEigenvalue}
\begin{aligned}
i\zeta = \pm i\left(j + \frac{N-1}{2} \right) , \; j = \frac{1}{2},\: \frac{3}{2},\: \frac{5}{2},\: \frac{7}{2},\: ....
\end{aligned}
\end{equation}
As such we have determined the eigenvalue for our eigenspinor-vectors for $N \geq 3$  and $N$ odd.

\subsubsection{N even}
We use the gamma matrices and spin connections as given in Eq.(\ref{EQ:GammaEven}) and Eq.(\ref{OmegaEven}) and set $\psi_{\mu} = (\psi_{\theta_{i}},\: \psi_{\theta_{N}})$, as such
\begin{equation}
\begin{aligned}
\psi_{\theta_{N}} =
\begin{pmatrix}
\psi_{\theta_{N}}^{(1)}\\
\psi_{\theta_{N}}^{(2)}\\
\end{pmatrix}
 , \; \psi_{\theta_{i}} =
\begin{pmatrix}
\psi_{\theta_{i}}^{(1)}\\
\psi_{\theta_{i}}^{(2)}\\
\end{pmatrix} .
\end{aligned}
\end{equation}
This allows us to rewrite Eq.(\ref{TTDirac}) and Eq.(\ref{Eq:TTcond.}) as
\begin{equation}\label{Eq:EvenGammaMuPsimu}
\begin{aligned}
\gamma^{\mu}\psi_{\mu} = 0 \implies \psi_{\theta_{N}}^{(1,2)} = \pm \frac{i}{\sin\theta_{N}}\widetilde{\gamma}^{\theta_{i}}\psi_{\theta_{i}}^{(1,2)}\: ,
\end{aligned}
\end{equation}
\begin{equation}
\begin{aligned}
\nabla^{\mu}\psi_{\mu} = 0 \implies \left(\partial_{\theta_{N}} + \left(N - \frac{1}{2} \right)\cot \theta_{N} \right)\psi_{\theta_{N}}^{(1,2)} = -\frac{1}{\sin^{2}\theta_{N}}\widetilde{\nabla}^{\theta_{i}}\psi_{\theta_{i}}^{(1,2)}\: ,
\end{aligned}
\end{equation}
\begin{equation}
\begin{aligned}
\gamma^{\mu}\nabla_{\mu}\psi_{\theta_{N}} = i \zeta \psi_{\theta_{N}} \implies \left(\partial_{\theta_{N}} + \left( \frac{N+1}{2}\right)\cot\theta_{N} \right)\psi^{(1,2)}_{\theta_{N}} \mp \frac{i}{\sin \theta_{N}}\widetilde{\gamma}^{\theta_{i}}\widetilde{\nabla}_{\theta_{i}}\psi_{\theta_{N}}^{(1,2)} = i \zeta \psi_{\theta_{N}}^{(2,1)}\: ,
\end{aligned}
\end{equation}
\begin{equation}\label{Eq:EvenGammaMuNablamu}
\begin{aligned}
\gamma^{\mu}\nabla_{\mu}\psi_{\theta_{i}} = i \zeta \psi_{\theta_{i}} \implies \left(\partial_{\theta_{N}} + \left(\frac{N-1}{2} \right)\cot\theta_{N} \right)\psi_{\theta_{i}}^{(1,2)}+ & 2\cot\theta_{N}\widetilde{\Sigma}_{\theta_{i} \: }^{\: \theta_{j}}\psi_{\theta_{j}}^{(1,2)}\\
 \mp & \frac{i}{\sin\theta_{N}}\widetilde{\gamma}^{\theta_{j}}\widetilde{\nabla}_{\theta_{j}}\psi_{\theta_{i}}^{(1,2)} = i \zeta \psi_{\theta_{i}}^{(2,1)}\: .
\end{aligned}
\end{equation}
We set
\begin{equation}
\begin{aligned}
\psi_{\theta_{N}}^{(1)} & = A^{(1)}\widetilde{\psi}_{(\lambda)} \: ; \: \psi_{\theta_{N}}^{(2)} = i A^{(2)}\widetilde{\psi}_{(\lambda)}\\
\psi_{\theta_{i}}^{(1)} & = C^{(1)}\widetilde{\nabla}_{\theta_{i}}\widetilde{\psi}_{(\lambda)} + D^{(1)}\widetilde{\gamma}_{\theta_{i}}\widetilde{\psi}_{(\lambda)} \: ; \: \psi_{\theta_{i}}^{(2)} = i C^{(2)}\widetilde{\nabla}_{\theta_{i}}\widetilde{\psi}_{(\lambda)} + i D^{(2)}\widetilde{\gamma}_{\theta_{i}}\widetilde{\psi}_{(\lambda)}\: .
\end{aligned}
\end{equation}
Substituting these into Eqs.(\ref{Eq:EvenGammaMuPsimu}) and (\ref{Eq:EvenGammaMuNablamu}) we have the same results as we did for the $N$ odd case. That is, we find that the eigenvalues are the same as those for the $N$ odd case. We note that $N > 2$, since as discussed earlier, there are no TT eigenmodes for the surface $S^{2}$.

\subsection{Spinor-vector TT-Modes II on $S^{N}$}
As discussed earlier in this section ``TT mode II" are only possible for $N \geq 4$. We start by letting the eigenspinor-vector $\psi_{\theta_{N}}=0$, and the ``TT mode" eigenspinor-vector on $S^{N-1}$ will still be an eigenspinor-vector on $S^{N}$ with suitable coefficients.

\subsubsection{N odd}
Setting spinor-vector $\psi_{\theta_{i}}$ as
\begin{equation}
\begin{aligned}
\psi_{\theta_{i}} = \left(\frac{1}{\sqrt{2}}B^{(1)}\left(1 + i \gamma^{N} \right) + \frac{1}{\sqrt{2}}B^{(2)}\left(1-i\gamma^{N} \right) \right)\widetilde{\psi}_{\theta_{i}}\: ,
\end{aligned}
\end{equation}
where $B^{(1)}$ and $B^{(2)}$ are functions of $\theta_{N}$ only. Substituting into Eq.(\ref{Eq:TTDirac1}) by setting $B^{(1,2)}=\left(\sin\theta_{N} \right)^{-\left(\frac{N-3}{2} \right)}\mathbbm{B^{(1,2)}}$, we have
\begin{equation}
\begin{aligned}
\mathbbm{B}^{(1,2)}=\left(\sin\frac{\theta_{N}}{2}\right)^{\left|\frac{1}{2}\pm \widetilde{\zeta}\right| + \frac{1}{2}}\left(\cos\frac{\theta_{N}}{2} \right)^{\left|\frac{1}{2} \mp \widetilde{\zeta} \right|+ \frac{1}{2}}P_{n}^{\left|\frac{1}{2}\pm \widetilde{\zeta}\right|,\left|\frac{1}{2} \mp \widetilde{\zeta} \right|}(\cos\theta_{N})\: ,
\end{aligned}
\end{equation}
 where $P_{n}^{\left|\frac{1}{2}\pm \widetilde{\zeta}\right|,\left|\frac{1}{2} \mp \widetilde{\zeta} \right|}(\cos\theta_{N})$ is again the Jacobi Polynomial. $\widetilde{\zeta}$ is the spinor-vector eigenvalue of the ``TT mode I" on $S^{N-1}$, which is given as $\widetilde{\zeta} = \widetilde{j} + \frac{N-2}{2}$. Such that the eigenvalues on ``TT mode II" are
\begin{equation} \label{Eq:TTEigenValues}
\begin{aligned}
i\zeta = & \pm i \left(n + \left|\widetilde{\zeta}  \right| + \frac{1}{2}\right)=\pm i \left(j + \frac{N-1}{2} \right) , \: \: j = \frac{1}{2},\frac{3}{2},\frac{5}{2},...
\end{aligned}
\end{equation}
which are the same as those of ``TT mode I" on $S^{N}$.
\subsubsection{N even}
Setting
\begin{equation}
\begin{aligned}
\psi_{\theta_{i}}^{(1)}=B^{(1)}\widetilde{\psi}_{\theta_{i}} \: ; \: \: \psi_{\theta_{i}}^{(2)} = i B^{(2)}\widetilde{\psi}_{\theta_{i}}\: ,
\end{aligned}
\end{equation}
and using Eqs.(\ref{Eq:EvenGammaMuPsimu}) to (\ref{Eq:EvenGammaMuNablamu}), we find that the eigenvalue is still given as Eq.(\ref{Eq:TTEigenValues}). Using the eigenvalues for our spinors and spinor-vectors we can determine our potential for spin-3/2 particles near Schwarzschild black holes.


\section{The radial equation and the potential function}\label{Sec:Potential}\setcounter{equation}{0}
\par In this section we are going to obtain the radial equation and the effective potential for the spin-3/2 field in the $D$-dimensional Schwarzschild black hole space-time. Since the mode function of the spin-3/2 field will be represented by the spinor-vector wave functions, we have to do the construction analogous to the details with the spinor and the vector fields. In the study of Maxwell fields it has been shown that there are two physical modes with different mode functions \cite{Lu2007, Lu2001}. One is related to the scalar spherical harmonics, and another one is related to the vector spherical harmonics, these are also known as the ``longitudinal" and ``transverse" parts of a vector field \cite{Rubin1985}. In our case there are ``non TT eigenmodes" and ``TT eigenmodes" on $S^{N}$, where we may obtain two physical modes related to these eigenmodes for our spin-3/2 field case.

\subsection{Massless Rarita-Schwinger field for $D$-Dimensions}

To begin we need to define our metric as
\begin{equation}
\begin{aligned}
ds^{2} = -fdt^{2} + \frac{1}{f}dr^{2} + r^{2}d\bar{\Omega}^{2}_{N}\: ,
\end{aligned}
\end{equation}
where $f = 1 - \left(\frac{2M}{r}\right)^{D-3}$ and $D=N+2$ is the dimensions of the space-time.  The $d\bar{\Omega}_{N}$ is the metric for the $N$ sphere, and we denote terms from the $N$ sphere with over bars. We will use the massless form of the Rarita-Schwinger equation to represent the spin-3/2 field,
\begin{equation}\label{RaritaSchwingerEq}
\gamma^{\mu \nu \alpha}\nabla_{\nu}\psi_{\alpha} = 0\: ,
\end{equation}
where the anti-symmetric Dirac gamma product is given as
\begin{equation}
\gamma^{\mu \nu \alpha} = \gamma^{[\mu}\gamma^{\nu}\gamma^{\alpha]} = \gamma^{\mu}\gamma^{\nu}\gamma^{\alpha} - \gamma^{\mu}g^{\nu \alpha} + \gamma^{\nu}g^{\mu \alpha} - \gamma^{\alpha}g^{\mu \nu}\: .\\
\end{equation}
We choose the following gamma matrices,
\begin{equation}\label{gammamatrices}
\begin{aligned}
\gamma^{0} &= i\sigma^{3} \otimes \mathbbm{1} &&\implies &&&\gamma^{t} = \frac{1}{\sqrt{f}}(i\sigma^{3} \otimes \mathbbm{1}) ,\\
\gamma^{i} &= \sigma^{1} \otimes \bar{\gamma}^{i} &&\implies &&&\gamma^{\theta_{i}} = \frac{1}{r}(\sigma^{1}\otimes \bar{\gamma}^{\theta_{i}}) ,\\
\gamma^{D-1} &= \sigma^{2} \otimes \mathbbm{1} &&\implies  &&&\gamma^{r} = \sqrt{f}(\sigma^{2} \otimes \mathbbm{1}) ,
\end{aligned}
\end{equation}
with $\mathbbm{1}$ being the $2^{\left(\frac{D-2}{2}\right)}\times2^{\left(\frac{D-2}{2}\right)}$ unit matrix for the case of $D$ even and the $2^{\left( \frac{D-3}{2} \right)}\times2^{\left(\frac{D-3}{2}\right)}$ unit matrix for the case of $D$ odd. $\sigma^{i} \: \left(i = 1 , 2 ,3\right)$ are the Pauli matrices and $\bar{\gamma}^{\theta_{i}}$ are the Dirac matrices for the $N$ sphere. The non-zero spin connections are
\begin{equation}\label{SpinConnections}
\begin{aligned}
\omega_{t} &= - \frac{f'}{4}(\sigma^{1} \otimes \mathbbm{1}) ,\\
\omega_{\theta_{i}} &= \mathbbm{1} \otimes \bar{\omega}_{\theta_{i}} + \frac{\sqrt{f}}{2}(i\sigma^{3} \otimes \bar{\gamma}_{\theta_{i}}) .
\end{aligned}
\end{equation}
And the non-zero triple gamma products are given as
\begin{equation}\label{TripleGamma}
\begin{aligned}
\gamma^{t\theta_{i}r} &= -\frac{1}{r}\left(\mathbbm{1} \otimes \bar{\gamma}^{\theta_{i}} \right), & \gamma^{t \theta_{i} \theta_{j}} &= \frac{1}{r^{2}\sqrt{f}}\left(i \sigma^{3} \otimes\bar{\gamma}^{\theta_{i}\theta_{j}}\right) ,\\
\gamma^{r\theta_{i}\theta_{j}} &= \frac{\sqrt{f}}{r^{2}}\left(\sigma^{2}\otimes\bar{\gamma}^{\theta_{i}\theta_{j}}\right) , & \gamma^{\theta_{i}\theta_{j}\theta_{k}} &= \frac{1}{r^{3}}\left(\sigma^{1}\otimes \bar{\gamma}^{\theta_{i}\theta_{j}\theta_{k}} \right) , \\
\end{aligned}
\end{equation}
where $\bar{\gamma}^{\theta_{i}\theta_{j}}=\bar{\gamma}^{\theta_{i}}\bar{\gamma}^{\theta_{j}} - \bar{g}^{\theta_{i}\theta_{j}}$ is the antisymmetric product of two Dirac matrices.

\subsection{With non-TT eigenfunctions}\label{Sec:SpinorPot}

\par We represent our radial, temporal and angular parts as $\psi_{r}$, $\psi_{t}$ and $\psi_{\theta_{i}}$. The radial and temporal parts will behave as spinors on $S^{N}$ and we write them as
\begin{equation}\label{PhiRANDPhiT}
\begin{aligned}
\psi_{r} = \phi_{r} \otimes \bar{\psi}_{(\lambda)} \: \: \textrm{and} \: \: \psi_{t} = \phi_{t} \otimes \bar{\psi}_{(\lambda)}, \\
\end{aligned}
\end{equation}
where $\bar{\psi}_{(\lambda)}$ is an eigenspinor on the $S^{N}$, with eigenvalues $i\bar{\lambda}$. The angular part, however, will behave as a spinor-vector on $S^{N}$ and can be written as Eq.(\ref{Eq:NonTTEigenVectors}). However, it is more convenient to write it as
\begin{equation}\label{AngularPhi}
\begin{aligned}
\psi_{\theta_{i}} = \phi^{(1)}_{\theta} \otimes \bar{\nabla}_{\theta_{i}}\bar{\psi}_{(\lambda)} + \phi^{(2)}_{\theta} \otimes \bar{\gamma}_{\theta_{i}}\bar{\psi}_{(\lambda)},\\
\end{aligned}
\end{equation}
where $\phi^{(1)}_{\theta}, \: \phi^{(2)}_{\theta}$ are functions of $r$ and $t$ which behave like 2-spinors. This is the same form as we have used for spinors when studying the 4-dimensional space-time \cite{chen2015gravitino}. Using Eq.(\ref{RaritaSchwingerEq}) we will derive our equations of motion and then try to rewrite them as Schr\"{o}dinger like equations. We will initially work in the Weyl gauge, where $\phi_{t} = 0$, to determine our equations of motion, and will then find a gauge invariant form.

\subsubsection{Equations of motion}
Firstly, consider the case where $\mu = t$ in Eq.(\ref{RaritaSchwingerEq}),
\begin{equation}\label{RaritaT}
\begin{aligned}
\gamma^{t \nu \alpha}\nabla_{\nu}\psi_{\alpha} =  0.
\end{aligned}
\end{equation}
By using Eq.(\ref{PhiRANDPhiT}) and Eq.(\ref{AngularPhi}), with the angular part separated, we have our first equation of motion in terms of $\phi_{r},\;\phi^{(1)}_{\theta}$ and $\phi^{(2)}_{\theta}$;
\begin{equation}\label{EoM1}
\begin{aligned}
0 = & -\left[i\bar{\lambda} + \frac{\sqrt{f}}{2}(D-2)(i\sigma^{3})\right]\phi_{r} + \left[i\bar{\lambda}\partial_{r} -\frac{1}{4}\frac{\left(D-2\right)\left(D-3\right)}{r\sqrt{f}}(i\sigma^{3}) + (D-3)\frac{i\bar{\lambda}}{2r}\right]\phi^{(1)}_{\theta}\\
&+\left[(D-2)\partial_{r} + \frac{i\bar{\lambda}(D-3)}{r\sqrt{f}}(i\sigma^{3}) + \frac{(D-2)(D-3)}{2r} \right]\phi^{(2)}_{\theta}\: .\\
\end{aligned}
\end{equation}
Next consider the case where $\mu = r$ in Eq.(\ref{RaritaSchwingerEq}),
\begin{equation}\label{RaritaR}
\begin{aligned}
\gamma^{r \nu \alpha}\nabla_{\nu}\psi_{\alpha} =  0.
\end{aligned}
\end{equation}
The second equation of motion is
\begin{equation}\label{EoM2}
\begin{aligned}
0 = & \left[-\frac{i\bar{\lambda}}{\sqrt{f}}\partial_{t}+\frac{i\bar{\lambda} f'}{4\sqrt{f}}\sigma^{1}-\frac{(D-3)(D-2)}{4r}\sigma^{2} +(D-3) \frac{i\bar{\lambda}\sqrt{f}}{2r}\sigma^{1} \right]\phi^{(1)}_{\theta} \\
&+ \left[-\frac{(D-2)}{\sqrt{f}}\partial_{t}+(D-2)\frac{f'}{4\sqrt{f}}\sigma^{1}+(D-3)\frac{i\bar{\lambda}}{r}\sigma^{2} + (D-2)(D-3)\frac{\sqrt{f}}{2r}\sigma^{1} \right]\phi^{(2)}_{\theta}\: .\\
\end{aligned}
\end{equation}
Finally for $\mu = \theta_{i}$,
\begin{equation}\label{RaritaTheta}
\begin{aligned}
\gamma^{\theta_{i} \nu \alpha}\nabla_{\nu}\psi_{\alpha} = 0.
\end{aligned}
\end{equation}
Giving us our final two equations of motion,
\begin{equation}\label{EoM3}
\begin{aligned}
0  = & \Bigg(\frac{1}{r\sqrt{f}}(i\sigma^{3})\partial_{t} + \frac{\sqrt{f}}{r}\sigma^{2}\partial_{r} +  \frac{f'}{4r\sqrt{f}}\sigma^{2}\\
&  +(D-4)\frac{\sqrt{f}}{2r^{2}}\sigma^{2}\Bigg)\phi^{(1)}- \left(\frac{D-4}{r^{2}}\sigma^{1} \right)\phi_{\theta}^{(2)} -\left(\frac{\sqrt{f}}{r}\sigma^{2}\right)\phi_{r}\:   \\
\end{aligned}
\end{equation}
and
\begin{equation}\label{EoM4}
\begin{aligned}
0 = & -\Bigg( \frac{i\bar{\lambda}}{r\sqrt{f}}(i\sigma^{3})\partial_{t} + \frac{i\bar{\lambda}\sqrt{f}}{r}\sigma^{2}\partial_{r} + \frac{i\bar{\lambda}f'}{4r\sqrt{f}}\sigma^{2}
 +\left(D-3\right)\left(D-4\right)\frac{1}{4r^{2}}\sigma^{1}\\
 & + (D-4)\frac{i\bar{\lambda}\sqrt{f}}{2r^{2}}\sigma^{2}\Bigg)\phi^{(1)}_{\theta} - \Bigg( \frac{D-3}{r\sqrt{f}}(i\sigma^{3})\partial_{t} + (D-3)\frac{\sqrt{f}}{r}\sigma^{2}\partial_{r} + (D-3)\frac{f'}{4r\sqrt{f}}\sigma^{2}\\
 & - (D-4)\frac{i\bar{\lambda}}{r^{2}}\sigma^{1}+ \left(D-3\right)\left(D-4\right)\frac{\sqrt{f}}{2r^{2}}\sigma^{2}\Bigg)\phi^{(2)}_{\theta}
  +\Bigg(\partial_{t} - \frac{(D-3)f}{2r}\sigma^{1} \\
  &- \frac{f'}{4}\sigma^{1}+ \frac{i\bar{\lambda}\sqrt{f}}{r}\sigma^{2} \Bigg)\phi_{r}\: .\\
\end{aligned}
\end{equation}
We now have our four equations of motion, Eqs.(\ref{EoM1}), (\ref{EoM2}), (\ref{EoM3}) and (\ref{EoM4}), in terms of $\phi_{r}$, $\phi_{\theta}^{(1)}$ and $\phi_{\theta_{2}}^{(2)}$. The functions $\phi_{r}$, $\phi_{\theta}^{(1)}$ and $\phi_{\theta}^{(2)}$ are not gauge invariant. In the next section we investigate the required gauge invariance and determine the appropriate transformations in order to create our gauge invariant radial equation.

\subsubsection{Gauge-invariant variable}\label{sec:Gauge}
If we consider a system where only gravitational forces are present then
\begin{equation}\label{GaugeSymmetry}
\gamma^{\mu\nu\alpha}\nabla_{\nu}\nabla_{\alpha}\varphi = \frac{1}{8}\gamma^{\mu\nu\alpha}R_{\nu\alpha\rho\sigma}\gamma^{\rho}\gamma^{\sigma}\varphi\: ,
\end{equation}
where $\varphi$ is a Dirac spinor. This allows our spinors-vectors to transform as
\begin{equation}\label{Eq:GaugeTransformation}
\psi'_{\mu} = \psi_{\mu} + \nabla_{\mu}\varphi\: ,
\end{equation}
and still have Eq.({\ref{RaritaSchwingerEq}) remain true, given that Eq.(\ref{GaugeSymmetry}) is equal to zero. This is not the case if our metric is charged, and we would need to introduce terms containing the electromagnetic field strength.
\par We can simplify the expression given in Eq.(\ref{GaugeSymmetry}) by exploiting the symmetry of the Riemann tensor,
\begin{equation}
\begin{aligned}
\gamma^{\mu}\gamma^{\nu}\gamma^{\alpha}\left(R_{\mu \nu \alpha \beta} + R_{\nu \alpha \mu \beta} + R_{\alpha \mu \nu \beta} \right) = 0  
\implies \gamma^{\mu}\gamma^{\nu}\gamma^{\alpha}R_{\mu \nu \alpha \beta} = - 2 \gamma^{\alpha}R_{\alpha \beta}\: .\\
\end{aligned}
\end{equation}
We also have that
\begin{equation}
\begin{aligned}
\gamma^{\mu}\gamma^{\nu}\gamma^{\alpha}\gamma^{\beta}R_{\mu \nu \alpha \beta} = - 2 R\: .
\end{aligned}
\end{equation}
Using these two identities Eq.(\ref{GaugeSymmetry}) becomes
\begin{equation}
\begin{aligned}
\frac{1}{8} \gamma^{\mu \nu \alpha} R_{\nu\alpha\rho \sigma} \gamma^{\rho}\gamma^{\sigma}  
= \frac{1}{4}\left(2 \gamma^{\alpha} R_{\alpha}^{\mu} - \gamma^{\mu} R \right)\varphi.\\
\end{aligned}
\end{equation}
This is zero for Ricci flat space-times like the $D$-dimensional Schwarzschild space-time. However, for de Sitter and anti-de Sitter space-times it does not vanish, so we would need to modify the covariant derivative in those cases in order to respect the gauge symmetry. This means we can perform the above transformation on our spinor-vector.
\par Firstly consider the transformation of $\phi_{r}$ and $\phi_{t}$. Take $\varphi = \phi \otimes \bar{\psi}_{(\lambda)}$, then Eq.(\ref{Eq:GaugeTransformation}) becomes
\begin{equation}
\begin{aligned}
\psi_{t}' = \psi_{t} + \nabla_{t}\psi & \implies \phi_{t}' = \phi_{t} + \partial_{r}\phi - \frac{f'}{4}\sigma^{1}\phi\: , \\
\psi_{r}' = \psi_{r} + \nabla_{r}\psi & \implies \phi_{r}' = \phi_{r} + \partial_{r}\phi\: .
\end{aligned}
\end{equation}
Next we consider the transformation of our angular components of $\psi_{\mu}$. They are given as
\begin{equation}
\begin{aligned}
& \psi_{\theta_{i}}' = \psi_{\theta_{i}} + \nabla_{\theta_{i}}\varphi,\\
\implies & \phi_{\theta}^{'(1)} \otimes \bar{\nabla}_{\theta_{i}}\bar{\psi}_{(\lambda)} + \phi_{\theta}^{'(2)}\otimes\bar{\gamma}_{\theta_{i}}\bar{\psi}_{(\lambda)} = \left(\phi_{\theta}^{(1)} + \phi \right)\otimes \bar{\nabla}_{\theta_{i}}\bar{\psi}_{(\lambda)} + \left(\phi_{\theta}^{(2)} + \frac{\sqrt{f}}{2}(i\sigma^{3})\phi \right)\otimes \bar{\gamma}_{\theta_{i}}\bar{\psi}_{(\lambda)}\: ,\\
\implies  & \phi_{\theta}^{'(1)} = \phi_{\theta}^{(1)} + \phi \: \:  ; \: \: \phi_{\theta}^{'(2)} = \phi_{\theta}^{(2)} + \frac{\sqrt{f}}{2}(i\sigma^{3})\phi\: .
\end{aligned}
\end{equation}
So clearly $\phi_{t}$, $\phi_{r}$, $\phi^{(1)}_{\theta}$ and $\phi^{(2)}_{\theta}$ are not gauge invariant. We need to perform a transformation of these spinors in order to obtain gauge invariant functions. We use the combination we have used in the $4$-dimensional space-time \cite{chen2015gravitino}
\begin{equation}
\begin{aligned}
\Phi = - \frac{\sqrt{f}}{2}(i\sigma^{3})\phi^{(1)}_{\theta} + \phi^{(2)}_{\theta}\: .
\end{aligned}
\end{equation}
Note that there is no dimensional dependence for our gauge invariant variable.

\subsubsection{Effective potential}
Using the gauge-invariant variable $\Phi$, Eq.(\ref{EoM1}), Eq.(\ref{EoM2}) and Eq.(\ref{EoM3}) become
\begin{equation}\label{GIEoM1}
\begin{aligned}
 \left((D-2)\partial_{r} + (D-3)\frac{i \bar{\lambda}}{r\sqrt{f}}(i\sigma^{3}) + \frac{(D-2)(D-3)}{2r} \right)\Phi &\\
+ \left(i\bar{\lambda} + \frac{D-2}{2}\sqrt{f}i\sigma^{3}\right)\partial_{r}\phi^{(1)}_{\theta} & = \left(i\bar{\lambda} + \frac{D-2}{2}\sqrt{f}(i\sigma^{3}) \right)\phi_{r}\: ,
\end{aligned}
\end{equation}
\begin{equation}\label{GIEoM2}
\begin{aligned}
 \left(-\frac{D-2}{\sqrt{f}}\partial_{t} + \frac{(D-2)(D-3)\sqrt{f}}{2r}\sigma^{1} + \frac{(D-2)f'}{4\sqrt{f}}\sigma^{1} + (D-3)\frac{i\bar{\lambda}}{r}\sigma^{2}  \right)\Phi &\\
 + \left( -\left(\frac{i\bar{\lambda}}{\sqrt{f}} + \frac{D-2}{2}i\sigma^{3} \right)\partial_{t} + \frac{i\bar{\lambda}f'}{4\sqrt{f}}\sigma^{1} - \frac{D-2}{8}f'\sigma^{2} \right)\phi^{(1)}_{\theta} &=0\: ,\\
\end{aligned}
\end{equation}
and
\begin{equation}\label{GIEoM3}
\begin{aligned}
\left(\frac{1}{\sqrt{f}}(i\sigma^{3})\partial_{t} + \sqrt{f}\sigma^{2}\partial_{r} + \frac{f'}{4\sqrt{f}}\sigma^{2} \right)\phi^{(1)}_{\theta} - \left(\frac{D-4}{r}\sigma^{1} \right)\Phi - \sqrt{f}\sigma^{2}\phi_{r} = 0\: .
\end{aligned}
\end{equation}
We have used that $f' = (D-3)(1-f)/r$ to simplify our equations. We can now use Eq.(\ref{GIEoM1}), Eq.(\ref{GIEoM2}) and Eq.(\ref{GIEoM3}) to derive a gauge invariant equation of motion in terms of only $\Phi$,

\begin{equation}\label{PHIEoM}
\begin{aligned}
0=&\left(\frac{D-2}{2}\sqrt{f} + \bar{\lambda}\sigma^{3} \right)\left[ -(D-2)\sigma^{1}\partial_{t} + \frac{D-2}{4}f' + \frac{D-3}{r}\sqrt{f}\left(\frac{D-2}{2}\sqrt{f}-\bar{\lambda}\sigma^{3} \right) \right]\Phi\\
 &-\left(\frac{D-2}{2}\sqrt{f} - \bar{\lambda}\sigma^{3} \right)\left[(D-2)f\partial_{r} + \frac{D-3}{r}\sqrt{f}\left(\frac{D-2}{2}\sqrt{f}-\bar{\lambda}\sigma^{3} \right) \right]\Phi\\
 &- \left(\frac{D-2}{2}\sqrt{f}- \bar{\lambda}\sigma^{3} \right)\left[\frac{D-4}{r}\sqrt{f}\left(\frac{D-2}{2}\sqrt{f} + \bar{\lambda}\sigma^{3} \right) \right]\Phi\: .\\
\end{aligned}
\end{equation}

\par Component wise $\Phi$ is given as

\begin{equation}
\Phi =
\begin{pmatrix}
\Phi_{1}(r)e^{-i \omega t} \\
\Phi_{2}(r)e^{-i \omega t}\\
\end{pmatrix} .
\end{equation}
Eq. (\ref{PHIEoM}) then becomes
\begin{equation}\label{FinalEoM1}
\begin{aligned}
\left(\frac{B}{A} \right)f\partial_{r}\Phi_{1} - \frac{f'}{4}\Phi_{1} - 2\left(\frac{B}{A}\right)\frac{D-3}{D-2}\frac{\sqrt{f}\bar{\lambda}}{r}\Phi_{1} + \frac{D-4}{D-2}\frac{\sqrt{f}}{r}B\Phi_{1} = i\omega\Phi_{2}\: ,
\end{aligned}
\end{equation}
\begin{equation}\label{FinalEoM2}
\begin{aligned}
\left(\frac{A}{B} \right)f\partial_{r}\Phi_{2} - \frac{f'}{4}\Phi_{2} + 2\left(\frac{A}{B}\right)\frac{D-3}{D-2}\frac{\sqrt{f}\bar{\lambda}}{r}\Phi_{2} + \frac{D-4}{D-2}\frac{\sqrt{f}}{r}A\Phi_{2} = i\omega\Phi_{1}\: ,
\end{aligned}
\end{equation}
where we have set
\begin{equation}
A = \frac{D-2}{2}\sqrt{f} + \bar{\lambda} \: \: \: \text{and} \: \: \: B =\frac{D-2}{2}\sqrt{f}-\bar{\lambda}\: .
\end{equation}
We can further simplify the above equation by defining
\begin{equation}
\widetilde{\Phi}_{1} = r^{\frac{D-4}{2}}\left(\frac{f^{1/4}}{\frac{D-2}{2}\sqrt{f} + \bar{\lambda}} \right)\Phi_{1} ;\:\:\:\: \widetilde{\Phi}_{2} = r^{\frac{D-4}{2}}\left(\frac{f^{1/4}}{\frac{D-2}{2}\sqrt{f}-\bar{\lambda}} \right)\Phi_{2}\: .
\end{equation}
Eq.(\ref{FinalEoM1}) and Eq.(\ref{FinalEoM2}) then become
\begin{equation}
\left(\frac{d}{dr_{*}} - W \right)\widetilde{\Phi}_{1} = i \omega \widetilde{\Phi}_{2} ; \:\:\: \left(\frac{d}{dr_{*}} + W\right)\widetilde{\Phi}_{2} = i\omega\widetilde{\Phi}_{1}\: ,
\end{equation}
where $r_{*}$ is the tortoise coordinate and is defined as $d/dr_{*} = f d/dr$. $W$ is known as the superpotential in supersymmetric quantum mechanics and is determined to be
\begin{equation}\label{EqW}
W = \frac{|\bar{\lambda}|\sqrt{f}}{r}\left( \frac{\left(\frac{2}{D-2} \right)^{2}|\bar{\lambda}|^{2} - 1- \frac{D-4}{D-2}\left(\frac{2M}{r} \right)^{D-3}}{\left(\frac{2}{D-2}\right)^{2}|\bar{\lambda}|^{2} - f}\right).
\end{equation}
This allows us to write our Schr\"{o}dinger like equation to describe our particles, where we have called this equation our radial equation, given as
\begin{equation}\label{Eq:RadialNonTT}
-\frac{d^{2}}{dr_{*}^{2}}\widetilde{\Phi}_{1} + V_{1}\widetilde{\Phi}_{1} = \omega^{2}\widetilde{\Phi}_{1} \: ; \: \:  -\frac{d^{2}}{dr_{*}^{2}}\widetilde{\Phi}_{2} + V_{2}\widetilde{\Phi}_{2} = \omega^{2}\widetilde{\Phi}_{2}\: ,
\end{equation}
with isospectral supersymmetric partner potentials \cite{Cooper1995}
\begin{equation}\label{Non-TTPotential}
V_{1,2} = \pm f \frac{dW}{dr} + W^{2}\: ,
\end{equation}
where $f = 1 - (2M/r)^{D-3}$ in $D$-dimensional space. As $\bar{\lambda} = n + (D-2)/2$ and $n = 0,1,2, ... $, we rewrite it as $\bar{\lambda} = j + (D-3)/2$, where $j = 1/2,3/2,5/2, ...$ such that $V(r)$ is explicitly given as
\begin{equation}
\begin{aligned}
V_{(1,2)} = & \frac{X\sqrt{f}(j + \frac{D-3}{2})}{r^{2}(X+Y)^{2}}\Bigg[X\left(\left(j + \frac{D-3}{2}\right)\sqrt{f} \pm \left(\frac{D-1}{2} \right)Y \mp 1 \right) \mp \left(\frac{2D^{2}-13D + 19}{D-2} \right)Y^{2} \Bigg]\\
 & +  \frac{(j + \frac{D-3}{2})\sqrt{f}Y^{2}}{r^{2}\left(X+Y \right)^{2}}\left(\frac{D-4}{D-2} \right)\Bigg[\left(j + \frac{D-3}{2} \right)\left(\frac{D-4}{D-2} \right)\sqrt{f}\pm 1 \mp \left( \frac{D-1}{2}\right)Y \Bigg] \\
 & +  \frac{X\sqrt{f}\left(j + \frac{D-3}{2} \right)}{r^{2}\left(X + Y \right)^{2}}\left[\pm 2 (D-4)Y \right].
 \label{EP}
\end{aligned}
\end{equation}
\begin{equation}
\begin{aligned}
X & = \left(\frac{2}{D-2} \right)^{2}\left(j -\frac{1}{2} \right)\left(j + \frac{2D-5}{2} \right)\:  ,\\
Y & = \left(\frac{2M}{r} \right)^{D-3}\: .
\end{aligned}
\end{equation}
Setting $D =4$ we find that our potential is the same as in Refs.\citep{chen2015gravitino,TC1990};
\begin{equation}
\begin{aligned}
V_{1,2} = & \frac{\left(j -\frac{1}{2} \right)(j + \frac{1}{2})\left(j + \frac{3}{2} \right)\sqrt{f}}{r^{2}(\left(j -\frac{1}{2} \right)\left(j + \frac{3}{2} \right)+\left(\frac{2M}{r} \right))^{2}}\\
& \times  \Bigg[\pm \frac{2M^{2}}{r^{2}} + \left(j -\frac{1}{2} \right)\left(j + \frac{3}{2} \right)\left(\left(j + \frac{1}{2}\right)\sqrt{f} \pm \left(\frac{3M}{r} \right) \mp 1 \right)  \Bigg] .\\
\end{aligned}
\end{equation}

\subsection{With the TT eigenfunctions}

\subsubsection{Equations of motion}
We set the radial and temporal parts $\psi_{r}$ and $\psi_{t}$, to be the same as the ``non-TT eigenfunctions" case given in Eq.(\ref{PhiRANDPhiT}). The angular part, $\psi_{\theta_{i}}$, can be written in terms of the TT mode eigenspinor-vector on $S^{N}$ as
\begin{equation}
\psi_{\theta_{i}} = \phi_{\theta} \otimes \bar{\psi}_{\theta_{i}} \:,
\end{equation}
where $\bar{\psi}_{\theta_{i}}$ is the TT mode eigenspinor-vector which includes the ``TT mode I" and ``TT mode II", and $\phi_{\theta}$ behaves like a 2-spinor. As we have done for the previous case, we will use the Weyl gauge and
consider the cases $\mu = t$, $\mu = r$, and $\mu = \theta_{i}$ for Eq.(\ref{RaritaSchwingerEq}). Applying the TT conditions on a sphere, namely $\bar{\gamma}^{\theta_{i}}\bar{\psi}_{\theta_{i}} =\bar{\nabla}^{\theta_{i}}\bar{\psi}_{\theta_{i}} = 0$, we have $\psi_{r} = 0$. We find that the equation of motion in this case is
\begin{equation}\label{TTEqofMo}
\left(i\sigma^{3}\partial_{t} + f\sigma^{2}\partial_{r} + \frac{f'}{4}\sigma^{2} + \frac{D-4}{2r}f\sigma^{2}+\frac{\sqrt{f}}{r}i\bar{\zeta}\sigma^{1} \right)\phi_{\theta} =0\: .
\end{equation}
In this case $\phi_{\theta}$ is gauge invariant, so we directly derive the radial equation in the next section.

\subsubsection{Effective potential}
Assuming $\phi_{\theta}$ is given as
\begin{equation}
\phi_{\theta} = \sigma^{2}
\begin{pmatrix}
\Psi_{\theta_{1}}e^{-i\omega t} \\
\Psi_{\theta_{2}}e^{-i\omega t} \\
\end{pmatrix} .
\end{equation}
Eq.(\ref{TTEqofMo}) can then be rewritten as
\begin{equation}\label{TTEqofMo2}
\begin{aligned}
&\left(f\partial_{r} + \frac{f'}{4} + \frac{D-4}{2r}f - \frac{\sqrt{f}}{r}\bar{\zeta} \right)\Psi_{\theta_{1}} = i\omega\Psi_{\theta_{2}}\: ,\\
&\left(f\partial_{r} + \frac{f'}{4} + \frac{D-4}{2r}f + \frac{\sqrt{f}}{r}\bar{\zeta}\right)\Psi_{\theta_{2}} = i\omega\Psi_{\theta_{1}}\: .\\
\end{aligned}
\end{equation}
These expressions can be simplified using the following transformations,
\begin{equation}
\bar{\Psi}_{\theta_{1}} = r^{\frac{D-4}{2}}f^{\frac{1}{4}}\Psi_{\theta_{1}}\: ,\: \: \bar{\Psi}_{\theta_{2}} = r^{\frac{D-4}{2}}f^{\frac{1}{4}}\Psi_{\theta_{2}}\: .
\end{equation}
We have the radial equations,
\begin{equation}\label{Eq:RadialTT}
-\frac{d^{2}}{dr_{*}^{2}}\bar{\Psi}_{\theta_{1}} + \mathbb{V}_{1}\bar{\Psi}_{\theta_{1}} = \omega^{2}\bar{\Psi}_{\theta_{1}} \: ; \: -\frac{d^{2}}{dr_{*}^{2}}\bar{\Psi}_{\theta_{2}} + \mathbb{V}_{2}\bar{\Psi}_{\theta_{2}}=\omega^{2}\bar{\Psi}_{\theta_{2}}\: ,
\end{equation}
where
\begin{equation}
\mathbb{V}_{1,2} = \pm f \frac{d\mathbb{W}}{dr} + \mathbb{W}^{2}\: ,
\end{equation}
and
\begin{equation}
\mathbb{W} = \frac{\sqrt{f}}{r}\bar{\zeta}\: .
\end{equation}
As $\bar{\zeta} = j + (D-3)/2$ where $j= 1/2, 3/2, 5/2, ...$. Our spinor-vector potentials are then explicitly given as
\begin{equation}\label{TTPotential}
\begin{aligned}
\mathbb{V}_{1,2} = & \pm\frac{\sqrt{1-\left(\frac{2M}{r}\right)^{D-3}}}{r^{2}}\left(j + \frac{D-3}{2} \right) \\
& \times \left[\frac{D-1}{2}\left(\frac{2M}{r} \right) - 1 \pm \sqrt{1 - \left(\frac{2M}{r} \right)^{D-3}}\left(j + \frac{D-3}{2} \right) \right] .
\end{aligned}
\end{equation}
This is the same potential as obtained in Ref.\cite{cho2007split}, where the radial equation of a spin-1/2 field on the general dimensional Schwarzschild black hole space-time is considered. We can say that the radial equation for the spin-3/2 field is equivalent to that of the spinor field case when the eigenmode on $S^{N}$ is the ``TT mode", with $\psi_{t} = \psi_{r} =0$, and only $\psi_{\theta_{i}}$ remains.


\section{Quasi normal modes}\label{Sec:QNMS}\setcounter{equation}{0}
In this section we focus on the QNMs for our ``non-TT eigenfunctions" spinor-vectors, where we will use the new potential that we have derived for the massless spin-3/2 fields. The potential that we have derived for the ``TT eigenfunctions" is the same as that seen for the spin-1/2 Dirac field, we therefore refer the reader to Ref.\cite{cho2007split} for the results of the ``TT eigenfunctions".
\subsection{Methods}
\par We have used two methods in order to determine the numerical values of our QNMs. We  have used the WKB method to 3rd and 6th order, and the improved AIM to calculate the numerical values of our QNMs. The 3rd order WKB method was developed by Iyer and Will \cite{iyer1987black} and the 6th order was developed by Konoplya \cite{konoplya2003quasinormal}.

\subsubsection{Implementation of the improved AIM}
 The improved AIM has been developed in the following papers \cite{cho2012new, cho2009asymptotic, doukas2009graviton, cho2010black}. In order to use this technique we must first perform a coordinate change so that we are operating on a compact space, we choose $\xi^{2} = 1 - 2M/r$. Our boundary conditions require that our particles are purely in-going at the horizon and purely out-going at infinity. Since our particles would exhibit plane wave behavior at these boundaries, we can write their wave functions as
\begin{equation}
\begin{aligned}
\widetilde{\Phi}_{1} & \sim e^{i \omega r_{*}} \: & \text{for} \: & r_{*} \rightarrow \infty\: ;\\
\widetilde{\Phi}_{1} & \sim e^{-i \omega r_{*}} \: & \text{for} \: & r_{*} \rightarrow -\infty\: . \\
\end{aligned}
\end{equation}
With $r_{*}$ as the tortoise coordinate, where the general formula for an $D$-dimensional tortoise coordinate is given in Ref. \cite{gibbons2002gravitational}
\begin{equation}\label{Eq:Rstar}
\begin{aligned}
r_{*} = r + \sum\limits_{n =1}^{D-3} \frac{e^{2\pi i\frac{n}{D-3}}}{D-3}2M \ln(r - 2Me^{2\pi i\frac{n}{D-3}})\: .
\end{aligned}
\end{equation}
Plugging Eq.(\ref{Eq:Rstar}) into our wave functions for our particles gives us our general behavior of the particles for our $D$ dimensional space
\begin{equation}
\begin{aligned}
\widetilde{\Phi}_{1} \sim e^{\pm \frac{2iM\omega}{1-\xi^{2}}} \prod_{n =1}^{D-3}\left(1-\left(1-\xi^{2} \right)\Theta(n) \right)^{\pm \frac{2iM\omega \Theta(n)}{D-3}}\left(\left(1-\xi^{2} \right)\Theta(n) \right)^{\pm \frac{2iM \omega \Theta(n)}{D-3}}\: ,
\end{aligned}
\end{equation}
where $\Theta(n)= e^{\frac{2 \pi i n }{D-3}}$. Clearly at the boundaries of our system, namely $\xi=1,0$, we would encounter asymptotic behavior. We extract this asymptotic behavior from $\phi$ and write
\begin{equation}
\begin{aligned}
\widetilde{\Phi}_{1} = \beta(\xi)\chi(\xi)\: ,
\end{aligned}
\end{equation}
where $\beta(\xi)$ contains our asymptotic behavior, and $\chi(\xi)$ satisfies the equation
\begin{equation}\label{Eq:AIM}
\begin{aligned}
\chi''(\xi) = \lambda_{0}\chi(\xi) + s_{0}\: .
\end{aligned}
\end{equation}
The functions $\lambda_{0}$ and $s_{0}$ are determined to be
\begin{equation}
\begin{aligned}
\lambda_{0} = - \left(2\frac{\beta'(\xi)}{\beta(\xi)} +  A\right) ,
\end{aligned}
\end{equation}
\begin{equation}
\begin{aligned}
s_{0} = -\left(\frac{\beta''(\xi)}{\beta(\xi)} + \frac{\beta'(\xi)}{\beta(\xi)}A + B \right) ,
\end{aligned}
\end{equation}
with
\begin{equation}
\begin{aligned}
A = \frac{\xi''}{\xi'} + \frac{f'}{f} \: ; \: \: B = \frac{1}{(f\xi')^{2}}\left(\omega^{2} - V\right) .
\end{aligned}
\end{equation}
We then apply the AIM to Eq.(\ref{Eq:AIM}) and after 200 iteration we obtain the results given in Tabs.(\ref{tab:45D})-(\ref{tab:89D}).

\subsection{Results}
\begin{figure}
\centering
\includegraphics[scale=1.25]{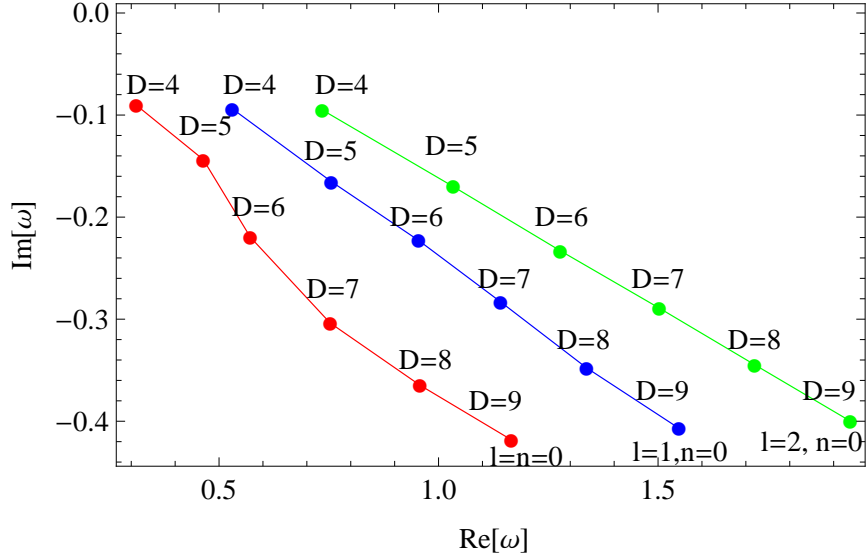}
\caption{Low lying QNMs of black hole space-times for D=4 to 9.}
\label{Fig:qnms}
\end{figure}
\par We present the results of our WKB and AIM calculation in Tabs. (\ref{tab:45D})-(\ref{tab:89D}). For fixed angular quantum number $l$ with specific dimension $D$, when the mode number $n$ gets larger, the real parts of the frequencies decrease and the imaginary parts, or the damping rates increase.  This indicates that the $n=0$ mode has the largest probability of being observed.  We also show the behavior of first few QNMs with various dimensions in Fig. (\ref{Fig:qnms}), which shows that both the frequency and the damping rate increase with the number of dimensions.
\par Comparing the WKB and AIM methods, we see that the WKB method returns results which are not as accurate as the AIM, however, the WKB is much easier to implement compared to the AIM. In order to obtain accurate results for QNMs using the AIM it is necessary for us to perform a large number of iterations of the AIM, which has the drawback of requiring large amounts of computation time. When $D=4,5,6$ we see that the AIM and WKB method are in strong agreement with each other, however, for higher values of $D$ we begin to see discrepancies between the AIM and WKB results. Due to these inconsistencies we have omitted some of the results for the 8 and 9 dimensional cases. These inconsistencies may be caused by the limitation of the WKB method when $n\sim l$ and/or when $l$ is large, in the higher dimensional cases.

\begin{table}[ht]
\caption{Low-lying ($n\leq l$, with $l=j-3/2$) spin-3/2 field quasi normal mode frequencies using the WKB and the AIM methods with D = 4,5.}
\centering
\begin{tabular}{| l | l || l | l || l || l | l | l | l || l | l || l | l || l | }
\hline
\multicolumn{5}{|c||}{4 Dimensions} & \multicolumn{5}{c|}{5 Dimensions}\\
\hline
$l$ & $n$ & 3rd order WKB & 6th order WKB & AIM & $l$ & $n$ & 3rd order WKB & 6th order WKB & AIM\\
\hline
0 & 0 & 0.3087-0.0902i &  0.3113-0.0902i &  0.3112-0.0902i & 0 & 0 & 0.4409-0.1529i &  0.4641-0.1436i &  0.4641-0.1435i \\
 \hline
1 & 0 & 0.5295-0.0938i &  0.5300-0.0938i &  0.5300-0.0937i & 1 & 0 & 0.7530-0.1653i &  0.7558-0.1652i &  0.7558-0.1651i \\
1 & 1 & 0.5103-0.2858i &  0.5114-0.2854i &  0.5113-0.2854i & 1 & 1 & 0.6902-0.5112i &  0.6989-0.5075i &  0.6988-0.5074i \\
\hline
2 & 0 & 0.7346-0.0949i &  0.7348-0.0949i &  0.7347-0.0948i & 2 & 0 & 1.0322-0.1700i &  1.0332-0.1700i &  1.0332-0.1700i \\
2 & 1 & 0.7206-0.2870i &  0.7210-0.2869i &  0.7210-0.2869i & 2 & 1 & 0.9869-0.5182i &  0.9900-0.5172i &  0.9899-0.5172i \\
2 & 2 & 0.6960-0.4844i &  0.6953-0.4855i &  0.6952-0.4855i & 2 & 2 & 0.9076-0.8835i &  0.9070-0.8868i &  0.9070-0.8868i \\
\hline
3 & 0 & 0.9343-0.0954i &  0.9344-0.0954i &  0.9343-0.0953i & 3 & 0 & 1.2998-0.1723i &  1.3003-0.1723i &  1.3003-0.1723i \\
3 & 1 & 0.9233-0.2876i &  0.9235-0.2876i &  0.9235-0.2875i & 3 & 1 & 1.2639-0.5221i &  1.2654-0.5217i &  1.2653-0.5216i \\
3 & 2 & 0.9031-0.4834i &  0.9026-0.4840i &  0.9025-0.4839i & 3 & 2 & 1.1985-0.8839i &  1.1974-0.8858i &  1.1974-0.8857i \\
3 & 3 & 0.8759-0.6835i &  0.8733-0.6870i &  0.8732-0.6870i & 3 & 3 & 1.1111-1.2586i &  1.1009-1.2748i &  1.1008-1.2748i \\
\hline
4 & 0 & 1.1315-0.0956i &  1.1315-0.0956i &  1.1315-0.0956i & 4 & 0 & 1.5617-0.1736i &  1.5620-0.1736i &  1.5619-0.1736i \\
4 & 1 & 1.1224-0.2879i &  1.1225-0.2879i &  1.1225-0.2879i & 4 & 1 & 1.5319-0.5244i &  1.5327-0.5242i &  1.5326-0.5242i \\
4 & 2 & 1.1053-0.4828i &  1.1050-0.4831i &  1.1049-0.4830i & 4 & 2 & 1.4761-0.8841i &  1.4751-0.8852i &  1.4750-0.8851i \\
4 & 3 & 1.0817-0.6812i &  1.0798-0.6830i &  1.0798-0.6829i & 4 & 3 & 1.3998-1.2546i &  1.3919-1.2639i &  1.3919-1.2638i \\
4 & 4 & 1.0530-0.8828i &  1.0485-0.8891i &  1.0484-0.8890i & 4 & 4 & 1.3070-1.6346i &  1.2874-1.6672i &  1.2873-1.6672i \\
\hline
5 & 0 & 1.3273-0.0958i &  1.3273-0.0958i &  1.3273-0.0957i & 5 & 0 & 1.8204-0.1744i &  1.8205-0.1744i &  1.8205-0.1744i \\
5 & 1 & 1.3196-0.2881i &  1.3196-0.2881i &  1.3196-0.2881i & 5 & 1 & 1.7947-0.5259i &  1.7952-0.5258i &  1.7951-0.5257i \\
5 & 2 & 1.3048-0.4824i &  1.3045-0.4826i &  1.3045-0.4825i & 5 & 2 & 1.7461-0.8842i &  1.7453-0.8848i &  1.7452-0.8848i \\
5 & 3 & 1.2839-0.6795i &  1.2826-0.6805i &  1.2826-0.6805i & 5 & 3 & 1.6783-1.2515i &  1.6724-1.2571i &  1.6723-1.2570i \\
5 & 4 & 1.2582-0.8794i &  1.2547-0.8832i &  1.2547-0.8831i & 5 & 4 & 1.5950-1.6274i &  1.5793-1.6478i &  1.5793-1.6478i \\
5 & 5 & 1.2284-1.0821i &  1.2221-1.0915i &  1.2220-1.0914i & 5 & 5 & 1.4983-2.0108i &  1.4696-2.0621i &  1.4695-2.0621i \\
\hline
\end{tabular}
\label{tab:45D}
\end{table}

\begin{table}[ht]
\caption{Low-lying ($n\leq l$, with $l=j-3/2$) spin-3/2 field quasinormal mode frequencies using the WKB and the AIM methods with D = 6,7.}
\begin{tabular}{| l | l || l | l || l || l | l | l | l || l | l || l | l || l | }
\hline
\multicolumn{5}{|c||}{6 Dimensions} & \multicolumn{5}{c|}{7 Dimensions}\\
\hline
$l$ & $n$ & 3rd order WKB & 6th order WKB & AIM & $l$ & $n$ & 3rd order WKB & 6th order WKB & AIM\\
\hline
0 & 0 & 0.5916-0.2260i &  0.5714-0.2197i &  0.5713-0.2197i & 0 & 0 & 0.7725-0.2978i &  0.7530-0.3037i &  0.7008-0.3036i \\
\hline
1 & 0 & 0.9479-0.2274i &  0.9548-0.2229i &  0.9547-0.2229i & 1 & 0 & 1.1441-0.2893i &  1.1415-0.2831i &  1.1231-0.2976i \\
1 & 1 & 0.8210-0.7101i &  0.8416-0.6761i &  0.8415-0.6761i & 1 & 1 & 0.9465-0.9065i &  0.9267-0.8783i &  0.9266-0.8782i \\
\hline
2 & 0 & 1.2745-0.2336i &  1.2771-0.2329i &  1.2771-0.2328i & 2 & 0 & 1.4998-0.2921i &  1.5026-0.2891i &  1.5026-0.2891i \\
2 & 1 & 1.1832-0.7155i &  1.1934-0.7085i &  1.1933-0.7084i & 2 & 1 & 1.3503-0.8967i &  1.3624-0.8752i &  1.3623-0.8752i \\
2 & 2 & 1.0199-1.2310i &  1.0220-1.2203i &  1.0220-1.2202i & 2 & 2 & 1.0742-1.5569i &  1.0498-1.5066i &  1.0498-1.5065i \\
\hline
3 & 0 & 1.5862-0.2376i &  1.5874-0.2373i &  1.5874-0.2373i & 3 & 0 & 1.8419-0.2960i &  1.8438-0.2949i &  1.8438-0.2949i \\
3 & 1 & 1.5140-0.7220i &  1.5187-0.7198i &  1.5187-0.7197i & 3 & 1 & 1.7229-0.9007i &  1.7323-0.8923i &  1.7322-0.8922i \\
3 & 2 & 1.3804-1.2296i &  1.3803-1.2284i &  1.3802-1.2283i & 3 & 2 & 1.4961-1.5419i &  1.4953-1.5198i &  1.4953-1.5198i \\
3 & 3 & 1.1990-1.7635i &  1.1734-1.7885i &  1.1734-1.7885i & 3 & 3 & 1.1835-2.2325i &  1.1146-2.2262i &  1.1146-2.2261i \\
\hline
4 & 0 & 1.8900-0.2400i &  1.8906-0.2399i &  1.8906-0.2399i & 4 & 0 & 2.1755-0.2991i &  2.1766-0.2986i &  2.1765-0.2986i \\
4 & 1 & 1.8298-0.7266i &  1.8323-0.7257i &  1.8323-0.7256i & 4 & 1 & 2.0760-0.9061i &  2.0818-0.9025i &  2.0817-0.9024i \\
4 & 2 & 1.7160-1.2302i &  1.7152-1.2305i &  1.7151-1.2304i & 4 & 2 & 1.8838-1.5389i &  1.8845-1.5300i &  1.8845-1.5300i \\
4 & 3 & 1.5584-1.7552i &  1.5396-1.7714i &  1.5396-1.7714i & 4 & 3 & 1.6124-2.2095i &  1.5732-2.2118i &  1.5731-2.2118i \\
4 & 4 & 1.3647-2.3005i &  1.3099-2.3698i &  1.3099-2.3698i & 4 & 4 & 1.2773-2.9207i &  1.1449-2.9964i &  1.1449-2.9964i \\
\hline
5 & 0 & 2.1890-0.2417i &  2.1894-0.2416i &  2.1894-0.2416i & 5 & 0 & 2.5035-0.3013i &  2.5042-0.3011i &  2.5041-0.3010i \\
5 & 1 & 2.1371-0.7299i &  2.1387-0.7294i &  2.1386-0.7293i & 5 & 1 & 2.4177-0.9105i &  2.4213-0.9087i &  2.4212-0.9087i \\
5 & 2 & 2.0378-1.2311i &  2.0369-1.2315i &  2.0368-1.2315i & 5 & 2 & 2.2504-1.5392i &  2.2507-1.5351i &  2.2506-1.5351i \\
5 & 3 & 1.8982-1.7499i &  1.8841-1.7603i &  1.8841-1.7603i & 5 & 3 & 2.0107-2.1974i &  1.9842-2.2013i &  1.9841-2.2013i \\
5 & 4 & 1.7246-2.2867i &  1.6825-2.3309i &  1.6825-2.3308i & 5 & 4 & 1.7101-2.8899i &  1.6167-2.9406i &  1.6166-2.9405i \\
5 & 5 & 1.5215-2.8396i &  1.4372-2.9604i &  1.4372-2.9604i & 5 & 5 & 1.3583-3.6159i &  1.1548-3.7942i &  1.1547-3.7941i \\
\hline
\end{tabular}
\label{tab:67D}
\end{table}

\begin{table}[ht]
\caption{Low-lying ($n\leq l$, with $l=j-3/2$) spin-3/2  field quasinormal mode frequencies using the WKB and the AIM methods with D = 8,9.}
\begin{tabular}{| l | l || l | l || l || l | l | l | l || l | l || l | l || l | }
\hline
\multicolumn{5}{|c||}{8 Dimensions} & \multicolumn{5}{c|}{9 Dimensions}\\
\hline
$l$ & $n$ &  3rd order WKB & 6th order WKB & AIM & $l$ & $n$ & 3rd order WKB & 6th order WKB & AIM\\
\hline
0 & 0 & 0.9675-0.3597i &  0.9577-0.3647i &  0.9675-0.3597i & 0 & 0 & 1.1706-0.4149i &  1.1654-0.4181i &  1.1706-0.4148i \\
\hline
1 & 0 & 1.3483-0.3498i &  1.3372-0.3477i &  1.3483-0.3498i & 1 & 0 & 1.5593-0.4062i &  1.5473-0.4069i &  1.5593-0.4062i \\
1 & 1 & 1.0776-1.0933i &  1.0370-1.0848i &  1.0775-1.0933i & 1 & 1 & 1.2078-1.2636i &  1.1700-1.2482i &  1.2078-1.2635i \\
\hline
2 & 0 & 1.7213-0.3485i &  1.7199-0.3446i &  1.7213-0.3484i & 2 & 0 & 1.9438-0.4026i &  1.9376-0.3998i &  1.9437-0.4026i \\
2 & 1 & 1.5065-1.0692i &  1.5036-1.0437i &  1.5064-1.0692i & 2 & 1 & 1.6581-1.2319i &  1.6427-1.2082i &  1.6580-1.2318i \\
2 & 2 & 1.0973-1.8731i &  1.0016-1.8170i &  1.0972-1.8730i & 2 & 2 & 1.0963-2.1769i &  0.9333-2.1004i &  1.0962-2.1768i \\
\hline
3 & 0 & 2.0845-0.3507i &  2.0856-0.3484i &  2.0844-0.3507i & 3 & 0 & 2.3216-0.4029i &  2.3202-0.4000i &  2.3215-0.4028i \\
3 & 1 & 1.9099-1.0665i &  1.9192-1.0496i &  1.9099-1.0664i & 3 & 1 & 2.0850-1.2224i &  2.0882-1.2003i &  2.0849-1.2223i \\
3 & 2 & 1.5665-1.8349i &  1.5456-1.7789i &  1.5665-1.8348i & 3 & 2 & 1.6038-2.1138i &  1.5438-2.0212i &  1.6037-2.1137i \\
3 & 3 & 1.0897-2.6898i &  0.9046-2.6246i &  1.0896-2.6897i & 3 & 3 & 0.9361-3.1458i &  0.5638-3.0044i &  0.9361-3.1457i \\
\hline
4 & 0 & 2.4398-0.3534i &  2.4410-0.3522i &  2.4398-0.3534i & 4 & 0 & 2.6931-0.4046i &  2.6934-0.4027i &  2.6930-0.4046i \\
4 & 1 & 2.2931-1.0700i &  2.3017-1.0611i &  2.2931-1.0699i & 4 & 1 & 2.4923-1.2226i &  2.5009-1.2076i &  2.4922-1.2225i \\
4 & 2 & 2.0011-1.8219i &  2.0000-1.7896i &  2.0010-1.8218i & 4 & 2 & 2.0795-2.0861i &  2.0671-2.0180i &  2.0795-2.0861i \\
4 & 3 & 1.5816-2.6380i &  1.4930-2.5837i &  1.5815-2.6380i  & \: & \: & \: & \: &  \: \\
\hline
5 & 0 & 2.7896-0.3558i &  2.7904-0.3552i &  2.7895-0.3557i & 5 & 0 & 3.0595-0.4066i &  3.0601-0.4055i &  3.0594-0.4066i \\
5 & 1 & 2.6627-1.0745i &  2.6690-1.0698i &  2.6627-1.0745i & 5 & 1 & 2.8851-1.2260i &  2.8934-1.2168i &  2.8850-1.2259i \\
5 & 2 & 2.4091-1.8186i &  2.4114-1.8008i &  2.4090-1.8185i & 5 & 2 & 2.5261-2.0759i &  2.5286-2.0323i &  2.5260-2.0759i \\
5 & 3 & 2.0378-2.6105i &  1.9886-2.5773i &  2.0377-2.6105i & 5 & 3 & 1.9901-3.0021i &  1.8953-2.8832i &  1.9901-3.0020i \\
\hline
\end{tabular}
\label{tab:89D}
\end{table}

\clearpage


\section{Absorption probabilities}\label{Sec:Absorption}\setcounter{equation}{0}
\par In this section we consider the absorption probabilities associated with the ``non-TT eigenfunctions". In Ref.\cite{cho2008emission} a similar analysis is done for the spin-1/2 field which is equivalent to our ``TT eigenfunction" case for the spin-3/2 field. The analytic study of field absorption probabilities near black holes was pioneered by Unruh in 1976 \cite{unruh1976absorption}. However, his method was only able to determine the absorption probabilities for low energy particles. So in order to determine the entire spectrum of absorption probabilities we need to use the WKB method. We will give a brief overview of the Unruh method to determine the form of our absorption probabilities, and then provide the absorption probabilities we calculate when using the WKB method.
\subsection{Unruh method}

\par To implement the Unruh method we must consider three regions around the black holes: The near region, where $f(r) \rightarrow 0$, the central region, where $V(r) \gg \omega$, and the far region where $f(r) \rightarrow 1$. Approximations are obtained for each of the regions and then coefficients are determined by comparing and evaluating the solutions at the boundaries. We will for convenience denote $V_{1}$ as $V$ in the following section and write either potential explicitly where ambiguity may occur. For the case with non TT eigenfunctions, we use $V_{1}$, given in Eq.(\ref{Non-TTPotential})

\subsubsection{Near region}
In the near region $f(r) \rightarrow 0$, so Eqs.(\ref{Eq:RadialNonTT}) become,
\begin{equation}\label{Eq:NearnonTT}
\left(\frac{d^{2}}{dr_{*}^{2}} + \omega^{2} \right)\widetilde{\Phi}_{I} = 0 ,
\end{equation}
with the ingoing boundary condition near the event horizon. The solution in this case becomes
\begin{equation}
\widetilde{\Phi}_{I} = A_{I}e^{-i\omega r_{*}}.
\end{equation}

\subsubsection{Central region}
In this region we have that $V(r) \gg \omega$ and hence Eq.(\ref{Eq:RadialNonTT}) becomes
\begin{equation}\label{EqH1}
\left(\frac{d}{dr_{*}} + W \right)\left(\frac{d}{dr_{*}} - W \right)\widetilde{\Phi}_{II} = 0\: .
\end{equation}
Defining $H$ as
\begin{equation}\label{EqH}
H = \left(\frac{d}{dr_{*}} - W \right)\widetilde{\Phi}_{II}\: .
\end{equation}
The solution of Eqs.(\ref{EqH1}) and (\ref{EqH}) is
\begin{equation}\label{SolH}
H = B_{II}\left(\frac{1 + \sqrt{f}}{1-\sqrt{f}} \right)^{\frac{j}{D-3}+ \frac{1}{2}}\left(\frac{\left(\frac{2}{D-2} \right)\left(j + \frac{D-3}{2} \right)- \sqrt{f}}{\left(\frac{2}{D-2} \right)\left(j + \frac{D-3}{2} \right) + \sqrt{f}} \right) ,
\end{equation}
where substituting Eq.(\ref{SolH}) into Eq.(\ref{EqH}), we have a first order differential equation with solution
\begin{equation}
\begin{aligned}
\widetilde{\Phi}_{II} = A_{II}\left(\frac{1 + \sqrt{f}}{1-\sqrt{f}}\right)^{\frac{j}{D-3} + \frac{1}{2}} \left(\frac{\left( \frac{2}{D-2}\right)\left(j + \frac{D-3}{2}\right)-\sqrt{f}}{\left(\frac{2}{D-2} \right)\left(j + \frac{D-3}{2}\right)+ \sqrt{f}} \right) + B_{II}\Psi\: .
\end{aligned}
\end{equation}
\begin{equation}
\begin{aligned}
\Psi = & \left(\frac{1 + \sqrt{f}}{1 -\sqrt{f}} \right)^{\frac{j}{D-3} + \frac{1}{2}}\left(\frac{\left( \frac{2}{D-2}\right)\left(j + \frac{D-3}{2}\right)-\sqrt{f}}{\left(\frac{2}{D-2} \right)\left(j + \frac{D-3}{2}\right)+ \sqrt{f}}\right)\\
 & \times \left[\int\limits^{r}\frac{1}{f}\left(\frac{1 - \sqrt{f}}{1 +\sqrt{f}} \right)^{\frac{2j}{D-3} + 1}\left(\frac{\left( \frac{2}{D-2}\right)\left(j + \frac{D-3}{2}\right)+\sqrt{f}}{\left(\frac{2}{D-2} \right)\left(j + \frac{D-3}{2}\right)- \sqrt{f}}\right)^{2}dr' \right] .
\end{aligned}
\end{equation}

\subsubsection{Far region}
In the far region $f(r) \rightarrow 1$, Eq.(\ref{Eq:RadialNonTT}) becomes
\begin{equation}\label{Eq:FarnonTT}
\begin{aligned}
\frac{d^{2}}{dr_{*}^{2}}\widetilde{\Phi}_{III} - \left[\frac{\left(\left(j + \frac{D-4}{2} \right)^{2} - \frac{1}{4} \right)}{r^{2}} - \omega^{2} \right]\widetilde{\Phi}_{III} = 0\: .
\end{aligned}
\end{equation}
In this region $r_{*} \sim r$, the solution can be expressed as a Bessel function
\begin{equation}
\begin{aligned}
\widetilde{\Phi}_{III} = A_{III}\sqrt{r}J_{j+\frac{D-4}{2}}(\omega r) + B_{III}\sqrt{r}N_{j+\frac{D-4}{2}}(\omega r)\: .
\end{aligned}
\end{equation}
We can set the incoming amplitude of our field $\widetilde{\Phi}_{III}$ at $r \rightarrow \infty$ to one. This gives us that
\begin{equation}
\begin{aligned}
A_{III} + iB_{III} = \sqrt{2\pi \omega}\: .
\end{aligned}
\end{equation}
Taking $r \rightarrow 1$ in the near region gives us that
\begin{equation}
\begin{aligned}
A_{I} = A_{II}, \: \: B_{II} = -i\omega A_{I}\: .
\end{aligned}
\end{equation}
Matching the solutions for regions II and III by taking $f=1-(1/r)^{D-3}$ and $r\rightarrow\infty$, we find that the absorption probability is given as
\begin{equation}
\begin{aligned}
\left| A_{j}(\omega)\right|^{2} = 4\pi C^{2} \omega^{2j+D-3}\left(1 + \pi C^{2} \omega^{2j+D-3} \right)^{-2} \approx 4 \pi C^{2}\omega^{2j+D-3}\: ,
\end{aligned}
\end{equation}
where
\begin{equation}
\begin{aligned}
C = \frac{1}{2^{\frac{D-1}{D-3}j +\frac{D-1}{2}}\Gamma(j+\frac{D-2}{2})}\left(\frac{j+ \frac{2D-5}{2}}{j - \frac{1}{2}}\right) ,
\end{aligned}
\end{equation}
with $\Gamma$ denoting the gamma function and $\omega$ less than 1. This can be checked by taking $f=1-2/r$ and $D=4$, then we obtain the solution for the 4-dimensional case we studied in Ref.\cite{chen2015gravitino}.

\subsection{WKB method}
When using with the WKB method it is more convenient to take $Q(x) = \omega^{2} - V$ such that Eqs.(\ref{Eq:RadialNonTT}) and (\ref{Eq:RadialTT}) become
\begin{equation}
\begin{aligned}
\left(\frac{d^{2}}{dr_{*}^{2}} + Q \right)\widetilde{\Phi}_{1} = 0\: .
\end{aligned}
\end{equation}
For low energy particles $\omega \ll V$, we can use the first order WKB approximation. The result for this absorption probability is given in Ref.\cite{cho2005wkb}
\begin{equation}\label{FirstOrderWKB}
\begin{aligned}
|A_{j}| = e^{\left[-2 \int\limits_{x_{1}}^{x_{2}}\frac{dx'}{f(x')}\sqrt{-Q(x')} \right]} ,
\end{aligned}
\end{equation}
where $x =\omega r$ with $x_{1}$ with $x_{2}$ being the turning points. That is, $Q(x_{1},x_{2}) = 0$ or $V_{x_{1},x_{2}}=\omega^{2}$, for a given energy $\omega$ and potential $V$. For particles of energy $\omega^{2} \sim V$ the formula of Eq.(\ref{FirstOrderWKB}) no longer converges and therefore has no solution. For this energy region we will need to use a higher order WKB approximation. We use the method developed by Iyer and Will \cite{iyer1987black}, the absorption probability is given as
\begin{equation}
\begin{aligned}
|A_{j}(\omega)|^{2} = \frac{1}{1+ e^{2S(\omega)}}\: ,
\end{aligned}
\end{equation}
where
\begin{equation}
\begin{aligned}
S(\omega)  = & \pi k^{1/2}\left[\frac{1}{2}z_{0}^{2} + \left(\frac{15}{64}b^{2}_{3} - \frac{3}{16}b_{4} \right)z_{0}^{4} \right]\\
& + \pi k ^{1/2}\left[\frac{1155}{2048}b_{3}^{4} - \frac{315}{256}b_{3}^{2}b_{4} + \frac{35}{128}b^{2}_{4} + \frac{35}{64}b_{3}b_{5} - \frac{5}{32}b_{6} \right]z_{0}^{6} + \pi k^{-1/2}\left[\frac{3}{16}b_{4} -\frac{7}{64}b_{3}^{2} \right]\\
& - \pi k^{-1/2}\left[\frac{1365}{2048}b_{3}^{4} - \frac{525}{256}b_{3}^{2}b_{4} + \frac{85}{128}b_{4}^{2} + \frac{95}{64}b_{3}b_{5} - \frac{25}{32}b_{6} \right]z_{0}^{2}\: ,
\end{aligned}
\end{equation}
where $z_{0}^{2},b_{n}$ and $k$ are defined by the components of the Taylor series expansion of $Q(r)$ near $r_{0}$,
\begin{equation}
\begin{aligned}
Q & = Q_{0} + \frac{1}{2}Q''_{0}z^{2} + \sum\limits_{n=3} \frac{1}{n!}\left( \frac{d^{n}Q}{dx^{n}}\right)_{0}z^{n}\\
& = k\left[z^{2} - z^{2}_{0} + \sum\limits_{n=3}b_{n}z^{n} \right]\: .
\end{aligned}
\end{equation}
That is,
\begin{equation}
\begin{aligned}
z & = r - r_{0} \: ; \: \: z_{0}^{2} \equiv -2\frac{Q_{0}}{Q_{0}''} \: ;\: \: k \equiv \frac{1}{2}Q_{0}'' \: ;\\
  b_{n} & \equiv \left(\frac{2}{n!Q_{0}''} \right)\left(\frac{d^{n}Q}{dr_{*}^{n}} \right)_{0} \: ; \: \: \frac{d}{dr_{*}} = \left(1 - \left(\frac{2M}{r} \right)^{D-3} \right)\frac{d}{dr}\: .
\end{aligned}
\end{equation}
where $0$ denotes the maximum $Q$ and the primes denote derivatives. 

\begin{figure}
\begin{subfigure}{0.48\textwidth}
\includegraphics[width=\textwidth]{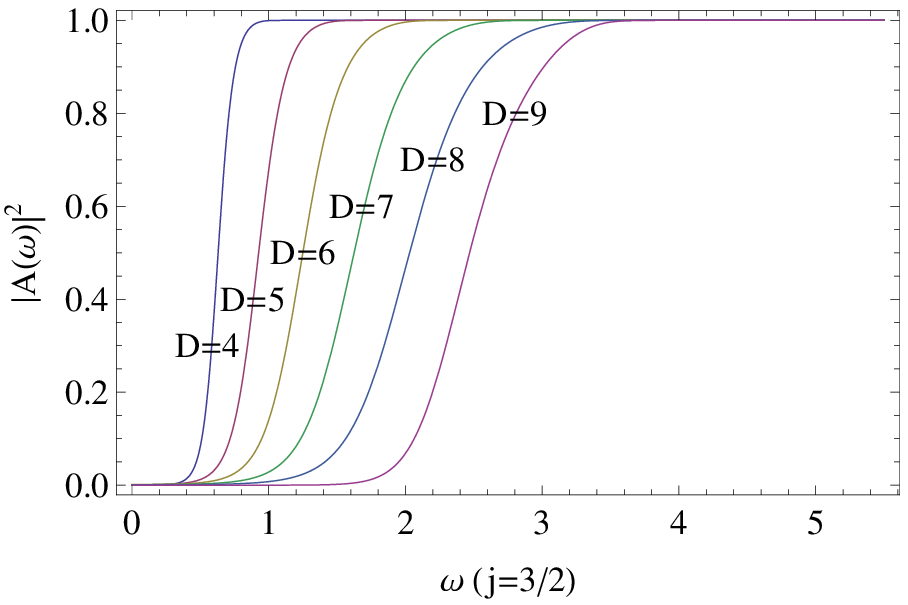}
\caption{$j=\frac{3}{2}$}
\label{Fig:AP15}
\end{subfigure}
\hfill
\begin{subfigure}{0.48\textwidth}
\includegraphics[width=\textwidth]{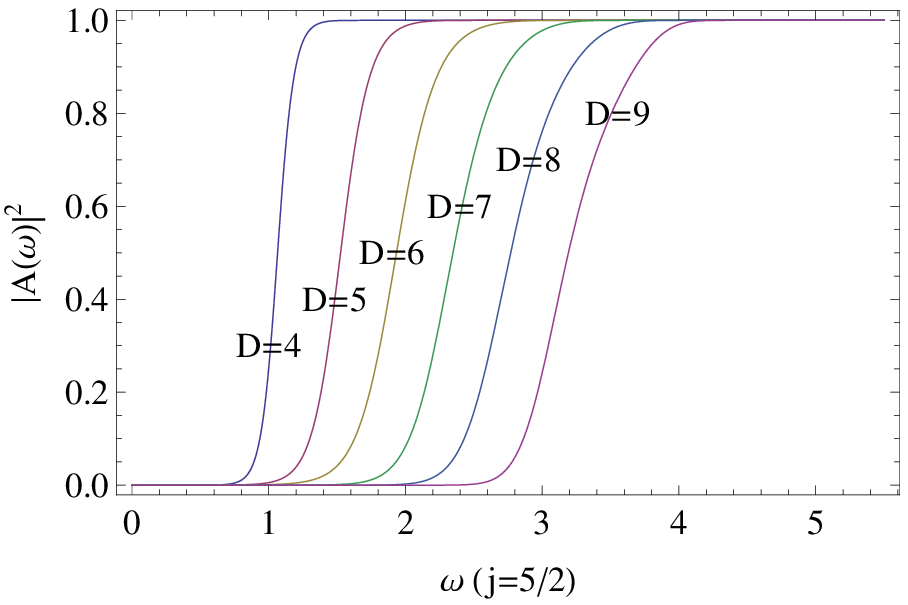}
\caption{$j=\frac{5}{2}$}
\label{Fig:AP25}
\end{subfigure}
\begin{subfigure}{0.48\textwidth}
\includegraphics[width=\textwidth]{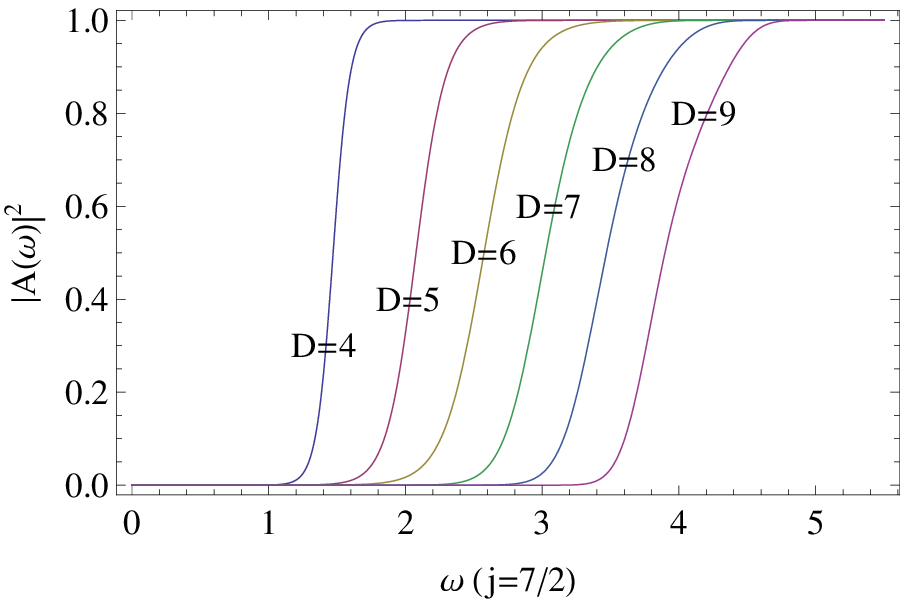}
\caption{$j=\frac{7}{2}$}
\label{Fig:AP25}
\end{subfigure}
\begin{subfigure}{0.48\textwidth}
\includegraphics[width=\textwidth]{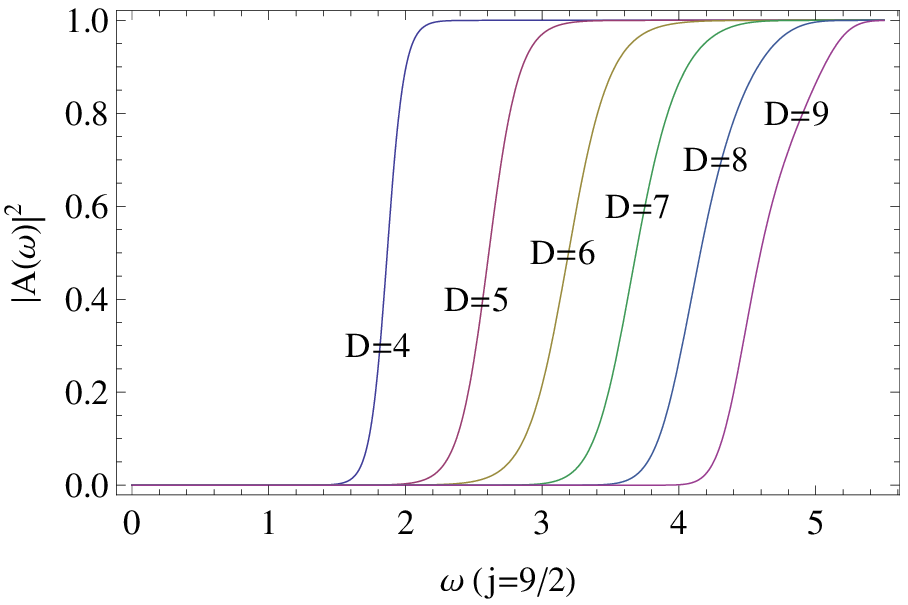}
\caption{$j=\frac{9}{2}$}
\label{Fig:AP25}
\end{subfigure}
\caption{spin-3/2 field absorption probabilities with various dimensions}
\label{Fig:APHD}
\end{figure}

\par In Fig.(\ref{Fig:APHD}) we can see that an increase in the value of $j$ results in an increase in the minimum required energy for total adsorption, we have observed and discussed this result in Ref.\cite{chen2015gravitino}. From Fig.(\ref{Fig:APHD}) we can clearly see that an increase in the number of dimensions results in an increase in the minimum required energy for total adsorption, similar to that seen for an increase in $j$. This occurs since, in both cases, our effective potential is getting larger and therefore the particles require more energy to tunnel through the effective potential.

\FloatBarrier


\section{Conclusion and discussion}\label{Sec:conclusions}\setcounter{equation}{0}

\par In this paper we have shown that by using the eigenvalues and eigenmodes of spinor-vectors on an $N$-sphere we can determine the effective potential for spin-3/2 fields in spherically symmetric space-times, with dimensions larger than 4. We have shown that there is a strong agreement between the potential that we have calculated and those calculated in other papers studying 4-dimensional space-times \cite{chen2015gravitino,cho2012new,cho2006asymptotic}. We have also investigated the QNMs for emitted fields from our $D$-dimensional black holes. Since the real part of our QNMs is a frequency we can see that the energy of emitted fields and the dimension of the space-time are directly related. This result is again seen when studying the absorption probabilities of particles near a Schwarzschild black hole. This suggests interesting results for the grey body factors of our $D$-dimensional black holes, where in order to make conclusions about the grey body factors we would need to study the cross sections of our black holes.
\par Our method requires that the space-time be spherically symmetrical which means we could use this method to study Reissner-Nordstr\"{o}m, AdS- and dS-black holes. In the case of the Reissner-Nordtr\"{o}m black holes it has been shown that for the extremal 4-dimensional black holes the asymptotic QNM frequencies are the same for spin-$0$, $1$, $3/2$ and $2$ \cite{cho2006asymptotic}. We would like to see if this is true for higher dimensional extremal Reissner-Nordstr\"{o}m black holes. In order to do this we must determine the covariant derivative related to the Reissner-Nordstr\"{o}m black hole space-time, where as stated earlier we must introduce terms with the Maxwell stress tensor and this derivative has been given in Refs. \cite{liu2014small}.
\par We can use our calculated potentials to determine the stability of the higher dimensional Schwarzschild black holes, a similar analysis is done in Refs. \cite{koda2003,koda2004}. The effective potential $V_{1}$ in Eq. (\ref{EP}) has a local minimum near the horizon when j=3/2 and $D=9$. For higher dimensions we see that this minimum becomes more negative when the number of dimensions are increased. We see the same thing occurs when j=5/2 and $D\geq14$, as can be seen in Fig. \ref{Fig:pd}. These behaviors are similar to the integer spin fields in some maximally symmetric space-times \cite{koda2003}. While the effective potential studied in this paper are all barrier like, the effective potentials of these higher dimensional space-times do warrant further studies.
\begin{figure}
\begin{subfigure}{0.48\textwidth}
\includegraphics[width=\textwidth]{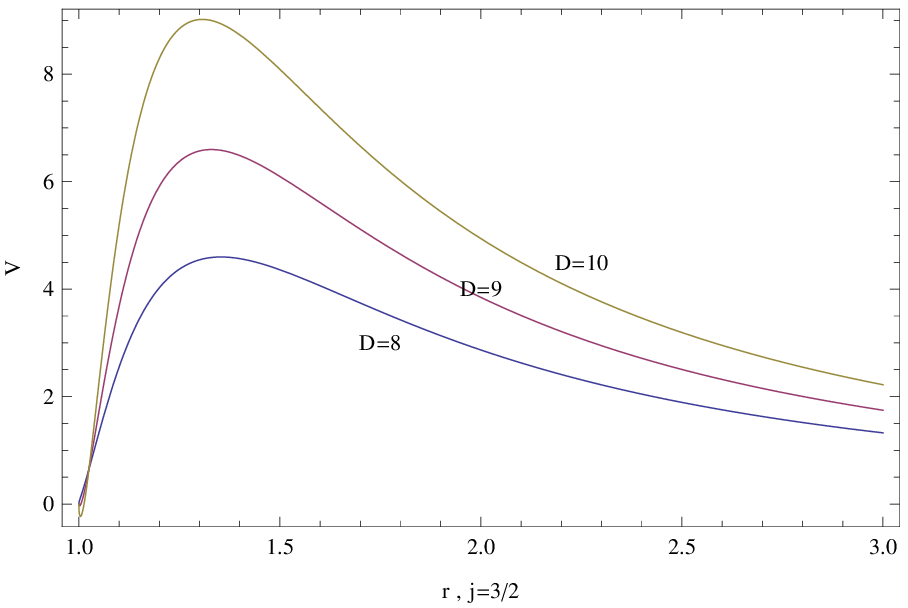}
\caption{$j=\frac{3}{2}$}
\label{Fig:pd1}
\end{subfigure}
\hfill
\begin{subfigure}{0.48\textwidth}
\includegraphics[width=\textwidth]{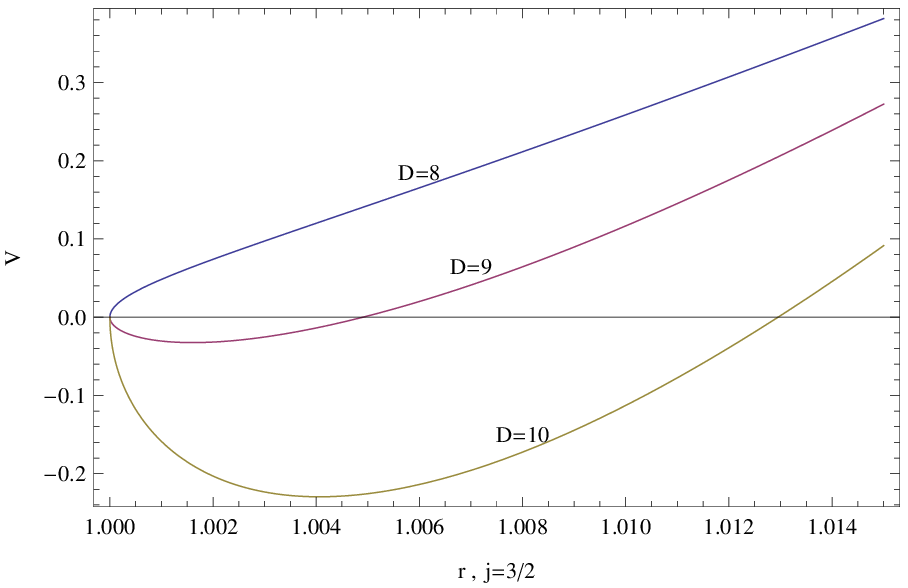}
\caption{$j=\frac{3}{2}$ when near the horizon.}
\label{Fig:pd2}
\end{subfigure}
\begin{subfigure}{0.48\textwidth}
\includegraphics[width=\textwidth]{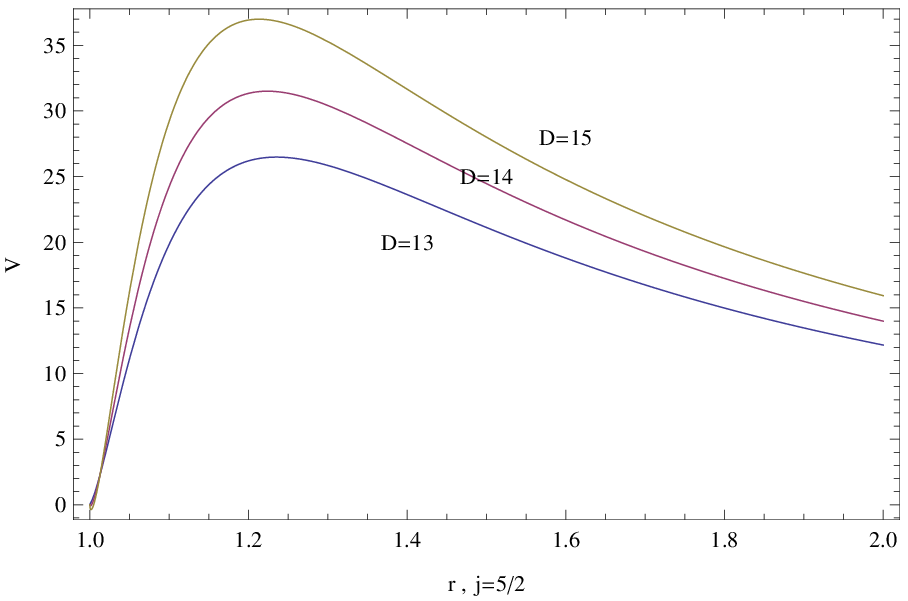}
\caption{$j=\frac{5}{2}$}
\label{Fig:pd3}
\end{subfigure}
\begin{subfigure}{0.48\textwidth}
\includegraphics[width=\textwidth]{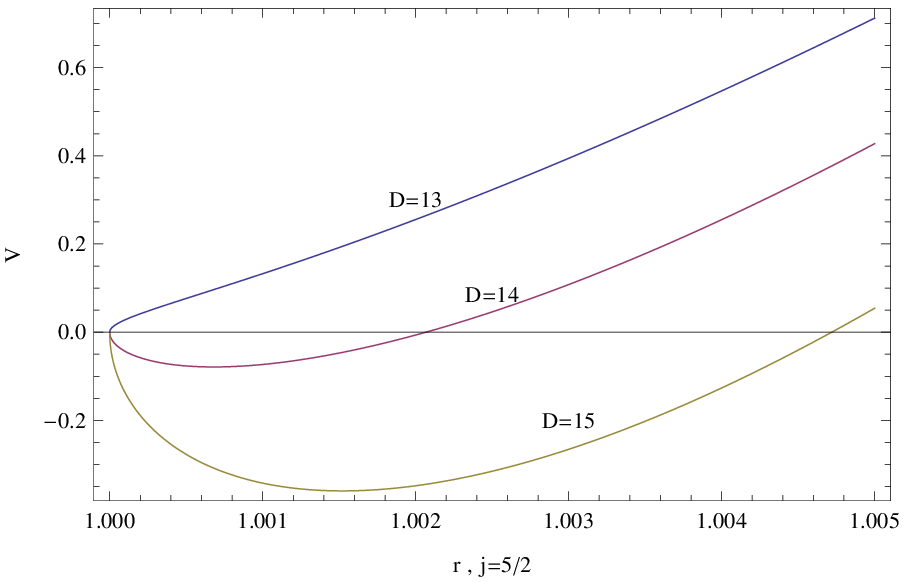}
\caption{$j=\frac{5}{2}$ when near the horizon.}
\label{Fig:pd4}
\end{subfigure}
\caption{spin-3/2 effective potential $V_{1}$ in higher dimensions.}
\label{Fig:pd}
\end{figure}


\acknowledgments

\noindent We would like thank Wade Naylor for his useful discussions during the production of this work. ASC and GEH are supported in part by the National Research Foundation of South Africa (Grant No: 91549). CHC and HTC are supported in part by the Ministry of Science and Technology, Taiwan, ROC under the Grant No. NSC102-2112-M-032-002-MY3. CHC especially thanks the National Research Foundation of South Africa (Grant No: 91549) for support while visiting the University of the Witwatersrand to finish this work.

\appendix
\renewcommand{\theequation}{\Alph{section}.\arabic{equation}}

\section{Eigen spinors on N-dimensional spheres}\label{Apend:NSphere}\setcounter{equation}{0}
The covariant derivative for our massless spinors is
\begin{equation}\label{CovariantSpinor}
\begin{aligned}
\nabla_{\mu}\psi = \partial_{\mu}\psi + \omega_{\mu}\psi,
\end{aligned}
\end{equation}
where
\begin{equation}
\begin{aligned}
&\omega_{\mu} = \frac{1}{2}\omega_{\mu a b}\Sigma^{a b}, \;  \Sigma^{a b}=\frac{1}{4}\left[\gamma^{a}, \gamma^{b} \right].
\end{aligned}
\end{equation}
In order to determine the eigenvalues for our spinors on $S^{N}$ we must consider the case of $N$ even and the case of $N$ odd. We begin with the  case of $N$ even.
\subsection{$N$ even}
In this case our gamma matrices are given as
\begin{equation}\label{EQ:GammaEven}
\begin{aligned}
&\gamma^{N} =
\begin{pmatrix}
0 & \mathbbm{1} \\
\mathbbm{1} & 0 \\
\end{pmatrix}
 , \: &
\gamma^{i} =
\begin{pmatrix}
0 & i\widetilde{\gamma}^{i} \\
-i\widetilde{\gamma}^{i} & 0 \\
\end{pmatrix} ,
\end{aligned}
\end{equation}
where $\mathbbm{1}$ is the $2^{\frac{N-2}{2}}\times2^{\frac{N-2}{2}}$ identity matrix. Our spin connections are then given as
\begin{equation}\label{OmegaEven}
\omega_{\theta_{i}}  =
\begin{pmatrix}
\widetilde{\omega}_{\theta_{i}} +  \frac{i}{2}\cos\theta_{N}\widetilde{\gamma}_{\theta_{i}}
& 0 \\
0 & \widetilde{\omega}_{\theta_{i}} - \frac{i}{2}\cos\theta_{N}\widetilde{\gamma}_{\theta_{i}}
 \\
\end{pmatrix}.
\end{equation}
Using the definition of the Dirac derivative given in Eq.(\ref{CovariantSpinor}) we have our spinor equation as
\begin{equation}\label{SpinorEq}
\begin{aligned}
 \gamma^{\mu}\nabla_{\mu}\psi_{(\lambda)}  
 = \left[\left(\partial_{\theta_{N}} + \frac{N-1}{2}\cot\theta_{N} \right)\gamma^{N} + \frac{i}{\sin\theta_{N}}
\begin{pmatrix}
0 & 1\\
-1 & 0 \\
\end{pmatrix}
\widetilde{\gamma}^{\theta_{i}}\widetilde{\nabla}_{\theta_{i}}\right]\psi_{(\lambda)} = i \lambda \psi_{(\lambda)}.
\end{aligned}
\end{equation}
We can choose to express $\psi_{(\lambda)}$ as
\begin{equation}\label{Psi(N-1)}
\begin{aligned}
\psi_{(\lambda)} = \begin{pmatrix}
\psi_{(\lambda)}^{(1)} \\
\psi_{(\lambda)}^{(2)} \\
\end{pmatrix}
=
\begin{pmatrix}
A_{(\lambda)}(\theta_{N})\widetilde{\psi}_{(\lambda)}\\
-iB_{(\lambda)}(\theta_{N})\widetilde{\psi}_{(\lambda)}\\
\end{pmatrix},
\end{aligned}
\end{equation}
where $\widetilde{\psi}_{(\lambda)}$ is the eigenspinor for the surface of $S^{N-1}$. Substituting Eq.(\ref{Psi(N-1)}) into Eq.(\ref{SpinorEq}) we find that
\begin{equation}\label{RelateAB}
\begin{aligned}
& \left(\partial_{\theta_{N}} + \frac{N-1}{2}\cot\theta_{N} + \frac{\widetilde{\lambda}}{\sin\theta_{N}} \right)A_{(\lambda)} = \lambda B_{(\lambda)}\: , \\
& \left(\partial_{\theta_{N}} + \frac{N-1}{2}\cot\theta_{N} - \frac{\widetilde{\lambda}}{\sin\theta_{N}} \right)B_{(\lambda)} = -\lambda A_{(\lambda)}\: .
\end{aligned}
\end{equation}
We can solve this by expressing $B_{\lambda}$ as a Jacobi polynomial
\begin{equation}
\begin{aligned}
B_{(\lambda)}(\theta_{N}) &= \left(\cos\frac{1}{2}\theta_{N} \right)^{l}\left(\sin\frac{1}{2}\theta_{N} \right)^{l+1}P_{n-l}^{((N/2)+l,(N/2)+l-1)}\left(\cos\theta_{N} \right)\\
&= (-1)^{n-l}A_{(\lambda)}(\pi - \theta_{N})\: .
\end{aligned}
\end{equation}
We require that $(n-l) \geq 0$ by restriction from our Jacobi polynomial  and the eigenvalue can then be written as
\begin{equation}\label{Eig(N)}
\begin{aligned}
i\lambda = \pm i \left(n + \frac{N}{2} \right), n = 0, 1, 2,....
\end{aligned}
\end{equation}

\subsection{$N$ odd}
In the case where $N$ is odd our gamma matrices are given as
\begin{equation}\label{EQ:GammaOdd}
\begin{aligned}
\gamma^{N} =
\begin{pmatrix}
\mathbbm{1} & 0\\
0 & \mathbbm{1}\\
\end{pmatrix} \:
; \: \: \gamma^{i} = \widetilde{\gamma}^{i},\\
\end{aligned}
\end{equation}
where $\mathbbm{1}$ is the identity matrix of size $2^{\frac{N-3}{2}}\times2^{\frac{N-3}{2}}$. The non-zero spin connection is determined to be
\begin{equation}
\begin{aligned}
\omega_{i} = \widetilde{\omega}_{\theta_{i}} - \frac{1}{2}\cos\theta_{N}\gamma^{N}\widetilde{\gamma}_{\theta_{i}} \:.
\end{aligned}
\end{equation}
Substituting the result for our spin connection into Eq.(\ref{CovariantSpinor}) we find that the spinor equation is
\begin{equation}
\begin{aligned}
 \gamma^{\mu}\nabla_{\mu}\psi_{(\lambda)}  & = \left[\left(\partial_{\theta_{N}} + \frac{N-1}{2}\cot\theta_{N} \right)\gamma^{N} + \frac{1}{\sin\theta_{N}}\widetilde{\gamma}^{\theta_{i}}\widetilde{\nabla}_{\theta_{i}}\right]\psi_{(\lambda)} = i \lambda \psi_{(\lambda)}\: .
\end{aligned}
\end{equation}
Choosing $\psi_{(\lambda)}$ as
\begin{equation}
\begin{aligned}
\psi_{(\lambda)} = \frac{1}{\sqrt{2}}\left(1 + i \gamma^{N} \right)A_{(\lambda)}(\theta_{N})\widetilde{\psi}_{(\lambda)} + \frac{1}{\sqrt{2}}\left(1 - i \gamma^{N} \right)B_{(\lambda)}(\theta_{N})\widetilde{\psi}_{(\lambda)}\: ,
\end{aligned}
\end{equation}
Which is the same relation between $A_{\lambda}$ and $B_{\lambda}$ as we had in Eq.(\ref{RelateAB}). Hence the eigenvalues will be the same as those for the case of $N$ even, given in Eq.(\ref{Eig(N)}).


\bibliography{NDimSpinPaper} 
\end{document}